\documentclass[a4paper]{article}
\usepackage{amsmath,amssymb,amsthm,amscd}
\def\be{\begin{equation}}
\def\lb{\label}

\DeclareMathOperator{\supp}{supp}

\DeclareMathOperator{\Ad}{Ad}

\DeclareMathOperator{\arctanh}{arctanh}

\DeclareMathOperator{\Ann}{ann}

\DeclareMathOperator{\id}{id} 
 
\DeclareMathOperator{\SO}{SO} \DeclareMathOperator{\Or}{O}
 
\DeclareMathOperator{\SU}{SU}

\newcommand{\D}{\bar D}
\newcommand{\p}{\overline{\mathcal{P}}}

\DeclareMathOperator{\const}{const}

\newcommand{\z}{{\bf z}}
\newcommand{\ii}{\mathbf{i}}\DeclareMathOperator{\pt}{pt}

\def\pd#1#2{\frac{\partial{#1}}{\partial{#2}}}
\def\pd1#1{\frac{\partial}{\partial#1}}
\def\d1#1{\frac{d}{d#1}}

\newcommand{\laplace}{\bigtriangleup}

\def\text#1{\mbox{#1}}

\newcommand{\bSth}{{\bf S}^3}
\newcommand{\bHth}{{\bf H}^3}

\newcommand{\bSn}{{\bf S}^n}
\newcommand{\bHn}{{\bf H}^n}

\newtheorem{Def}{Definition}\newtheorem{theore}{Theorem}
\newtheorem{Rem}{Remark}
\newtheorem{corollary}{Corollary}

\begin{document}

\author{A.V.~Shchepetilov\footnote{Department of Physics, Moscow State
University, 119992 Moscow, Russia, e-mail address:
alexey@quant.phys.msu.su}, I.E.~Stepanova\footnote{Institute for
Physics of Earth, RAS, bol.\ Gruzinskaja 10, 123995 Moscow,
Russia}}
\title{Two-body problem on spaces of constant curvature}
\date{}\maketitle
\begin{abstract}
The two-body problem with a central interaction on simply
connected constant curvature spaces of an arbitrary dimension is
considered. The explicit expression for the quantum two-body
Hamiltonian via a radial differential operator and generators of
the isometry group is found. We construct a self-adjoint extension
of this Hamiltonian. Some its exact spectral series are calculated
for several potential in the space ${\bf S}^3$.

We describe also the reduced classical mechanical system on a
homogeneous space of a Lie group in terms of the coadjoint action
of this group. Using this approach the description of the reduced
classical two-body problem on constant curvature spaces is given.
\vskip 20pt

\end{abstract}

\section{Introduction}

The simply connected constant curvature spaces ${\bf S}^n$ and
${\bf H}^n$ posses isometry groups as wide as the isometry group
for the Euclidean space ${\bf E}^n$ and have no selected points or
directions \cite{Wolf}. The one-body classical and quantum
problems in the central field on these spaces were studied in many
papers, among which we indicate basic ones \cite{Lip}-\cite{BKO}.

In contrast to the Euclidean case, the configuration spaces ${\bf
S}^n\times {\bf S}^n,{\bf H}^n\times {\bf H}^n$ of the two-body
problem on spaces ${\bf S}^n$ and ${\bf H}^n$ are not spaces of a
constant curvature. Only space isometries that preserve an
interaction potential enter in the symmetry group of such problem
a priori. This group does not suffice  to ensure the integrability
of the two-body problem. At the same time, no "hidden" symmetries
or other integrability tools are known for nontrivial potential.
Numerical experiments for the reduced classical two-body problem
in spaces ${\bf S}^n$ and ${\bf H}^n,\,n=2,3$ \cite{Shch991} show
the soft chaos in this system for some natural interactive
potentials. Numerical experiments (\cite{Cher}) and analytical
results (\cite{Zi}-\cite{MacPrz}) for the restricted classical
two-body problem on ${\bf S}^2$ and ${\bf H}^2$ also prove its
nonintegrability.

The classical mechanical two-body problem in constant curvature
spaces was first considered in \cite{Shch1}, where the method of
the Hamiltonian reduction of systems with symmetries \cite{MW} was
used to exclude the motion of a system as a whole. The description
of reduced mechanical systems, their classification, and
conditions for existence of a global dynamic were obtained using
explicit analytic coordinate calculations on a computer. In
\cite{Shch2}, an analogous quantum mechanical system was
considered in the two-dimensional case, i.e.\ on the spaces  ${\bf
S}^2$ ¨ ${\bf H}^2$. There, the quantum mechanical two-body
Hamiltonian was expressed through isometry group generators and a
radial differential operator. The structure of this expression is
similar to the structure of reduced Hamilton function. The idea
arises to seek a general procedure for simultaneous simplification
both classical and quantum problems without performing cumbersome
calculations. We present such a procedure in the present paper.
The obtained expression for the quantum two-body Hamiltonian is
useful for solving at least three problems.

First it enables us to prove that the two-body Hamiltonian with
the proper domain is self-adjoint. Secondly, using this
expression, one can reduce the spectral problem for the two-body
Hamiltonian to a sequence of systems of ordinary differential
equations enumerated by irreducible representations of the
isometry group. In the case of the sphere ${\bf S}^3$ we found all
separate differential equations for spectral values, which can be
solved in an explicit form for some interaction potentials. Thus,
although the two-body problem on spaces ${\bf S}^n$ and ${\bf
H}^n$ seems to be non-integrable in any sense, some its energy
levels can be explicitly found. Such a situation appears in so
called quasi exactly solvable models \cite{Us1}-\cite{Ush}. There
are two essential differences however. First, usually quasi
exactly solvable models are artificially constructed. Second exact
energy levels are obtained, as a rule, from one differential
equation with the specially selected potential, but this equation
has also other unknown spectral values. Conversely, in the problem
under consideration we select a separate differential equation
from systems of ordinary differential equations and find all its
spectral values.

Finally, from the obtained expression for the two-body Hamiltonian
we derive the Hamilton function of the reduced two-body classical
mechanical system, using the description of a reduced classical
mechanical system on a homogeneous space in terms of coadjoint
orbits of the corresponding Lie group, founded in section
\ref{ClassReduc}.

This paper is an essential revision of papers \cite{Sh001} and
\cite{ShchStep}, made by the first author. Some results in those
papers were not properly grounded. This was later done in
\cite{Shchep023}. Proofs in the present paper are extended and in
some technically difficult cases they contain references to
\cite{Shchep023}. More serious revision concerns the theorem
\ref{eigenvectorsS3} of the present paper. Because of a confusion
between left and right shifts on a group, the statement in
\cite{ShchStep} corresponding to this theorem contains some
additional erroneous eigenvectors.

\section{Notations}

Consider the sphere ${\bf S^n} $ as the space
$\mathbb{R}^n\cup\{\infty\}$ with the metric
\begin{equation}\lb{met_s} g_s=\left.\left(4R^2
\sum\limits_{i=1}^{n}dx_i^2\right)\right/
\left(1+\sum\limits_{i=1}^{n}x_i^2\right)^2,
\end{equation}
where $x_i,\quad i=1,\dots,n$ are Cartesian coordinates in
$\mathbb{R}^n$ and $R$ is the curvature radius. Let
$\rho^s(\cdot,\cdot)$ denotes the distance between two points in
${\bf S}^n$. The identity component of a whole isometry group for
${\bf S}^n$, acting from the left, is $\SO(n+1)$. The set
\begin{align}\begin{split}
X^s_{ij}&=x_i\pd1{x_j}-x_j\pd1{x_i}, \quad 1\leqslant i<j\leqslant
n,
\\ Y^s_i&=\frac12\left(1-\sum\limits_{j=1}^n
x_j^2 \right)\pd1{x_i} + x_i\sum\limits_{j=1}^n x_j\pd1{x_j},\quad
i=1,\dots,n,\end{split}\lb{basis_s}
\end{align}
is a base of Killing vector fields on ${\bf S}^n$. It corresponds
to some base in the Lie algebra $\mathfrak{so}(n+1)$.

Consider the hyperbolic space ${\bf H}^n$ as the unit ball
$D^n\subset\mathbb{R}^n$ with the metric
\begin{equation}\lb{met_h}
g_h=\left.\left(4R^2 \sum\limits_{i=1}^{n}dx_i^2\right)\right/
\left( 1-\sum\limits_{i=1}^{n}x_i^2\right)^2,\qquad
\sum\limits_{i=1}^{n} x_i^2<1. \end{equation}

Denote the distance between two points in ${\bf H}^n$ as
$\rho^h(\cdot,\cdot)$. Let $\Or_{0}(1,n)$ be an identity component
of a whole isometry group for ${\bf H}^n$, acting from the left.
Its Lie algebra is $\mathfrak{so}(1,n)$. The set
\begin{align}\begin{split}
X^h_{ij}&=x_i\pd1{x_j}-x_j\pd1{x_i}, \quad 1\leqslant i<j\leqslant
n, \\
Y^h_i&=\frac12\left(1+\sum\limits_{j=1}^n x_j^2
\right)\pd1{x_i} - x_i\sum\limits_{j=1}^n x_j\pd1{x_j},\quad
i=1,\dots,n
\end{split}\lb{basis_h}\end{align}
is a base of Killing vector fields on ${\bf H}^n$.

\section{Special forms of free Hamiltonians}

Let $Q_s={\bf S}^n\times{\bf S}^n$ and $Q_h={\bf H}^n\times{\bf
H}^n$ be the configuration spaces of the two-body problems in
${\bf S}^n$ and ${\bf H}^n$. The corresponding Hamiltonians are
defined as
\begin{equation}\lb{Ham_quan}
\widehat{H}_{s,h}=-\frac1{2m_1}\laplace_1-\frac1{2m_2}\laplace_2+
U(\rho^{s,h})\equiv\widehat{H}_0^{s,h}+U(\rho^{s,h}),
\end{equation}
where $\laplace_1$ and $\laplace_2$ are Laplace-Beltrami operators
in the direct factors of ${\bf S}^n\times{\bf S}^n$ and ${\bf
H}^n\times{\bf H}^n$, corresponding to the first and the second
particles, and $U$ is a central potential. Here and below the
subscript $"s"$ corresponds to the spherical case and the
subscript $"h"$ corresponds to the hyperbolic case.

According to the general concept of quantum mechanics \cite{SR}, a
domain of the operator $\widehat{H}_{s,h}$ must be a proper
everywhere dense subspace in the space
$\mathcal{L}^2\left(Q_{s,h},d\mu_{s,h}\right)$ of all square
integrable functions on $Q_{s,h}$. This subspace is chosen in such
a way that the operator $\widehat{H}_{s,h}$ becomes self-adjoint;
the corresponding measure $d\mu_{s,h}$ is the product of measures
on the direct factors of ${\bf S}^n\times{\bf S}^n$ $({\bf
H}^n\times{\bf H}^n)$, invariant w.r.t.\ the group $\SO(n+1)$
($\Or_{0}(1,n)$).

To express the total Hamiltonian $\widehat{H}_{s,h}$ through the
radial differential operator and generators of the isometry group,
it is suffices to find such an expression for the free
Hamiltonian. Recall (see, for example, \cite{DonGar}) that the
Laplace-Beltrami operator $\laplace$ on spaces ${\bf S}^n$ and
${\bf H}^n$ is respectively self-adjoint with domains
\begin{align*}
W^{2,2}_{s}&:=\left\{\phi\in\mathcal{L}^2({\bf
S}^n,d\mu_s)|\laplace\phi\in \mathcal{L}^2({\bf
S}^n,d\mu_s)\right\}, \\
W^{2,2}_{h}&:=\left\{\phi\in\mathcal{L}^2({\bf
H}^n,d\mu_h)|\laplace\phi\in \mathcal{L}^2({\bf
H}^n,d\mu_h)\right\}.
\end{align*}
The action of the
operator $\laplace$ is considered in the sense of distributions.
The operator $\laplace$ on ${\bf S}^n $ is essentially
self-adjoint on the space ${\bf C}^{\infty}({\bf S}^n)$ of smooth
functions, and the operator $\laplace$ on ${\bf H}^n $ is
essentially self-adjoint on the space of finite  smooth functions
${\bf C}^{\infty}_0({\bf H}^n)$ \cite{Str83}. Hence the free
Hamiltonian $\widehat{H}^{s,h}_0$ is self-adjoint on the product
\begin{equation}\label{DomSelfAdjointness}
W_{s,h}:=W^{2,2}_{s,h}\otimes W^{2,2}_{s,h}
\end{equation}
of two copies of the space $W^{2,2}_{s,h}$ corresponding
respectively to the first and the second particles.

Let $F^{s,h}_r$ be submanifolds of the space $Q_{s,h}$,
corresponding to the constant value $r$ of the function
$\tan(\rho^s/(2R))$ for the space $Q_s$ and the function
$\tanh(\rho^h/(2R))$ for the space $Q_h$. The submanifolds
$F^s_0,\;F^s_{\infty}$ are diffeomorphic to ${\bf S}^n$ (the value
$r=\infty$ corresponds to two diametrically opposite points of the
sphere ${\bf S}^n$) and $F^h_0$ is diffeomorphic to ${\bf H}^n$.
For $0<r<\infty$ the submanifold $F_r^s$ is a homogeneous
Riemannian space of the group $\SO(n+1)$ with the stationary
subgroup $K\cong\SO(n-1)$. For $0<r<1$ the submanifold $F_r^h$ is
a homogeneous Riemannian space of the group $\Or_{0}(1,n)$ with
the stationary subgroup $K$.

Up to a zero measure set it holds
$Q_s=\mathbb{R}_+\times(\SO(n+1)/K)$, where
$\mathbb{R}_+=(0,\infty)$ and also $Q_h=I\times(\Or_{0}(1,n)/K)$,
where $I=(0,1)$. The operators $-\widehat{H}_0^{s,h}$ are
Laplace-Beltrami ones for the metric $\widetilde{g}_{s,h}=2
m_1g^{(1)}_{s,h}+2m_2g^{(2)}_{s,h}$ on $Q_{s,h}$, where the
metrics $g^{(1)}_{s,h}$ and $g^{(2)}_{s,h}$ have either form
(\ref{met_s}) or (\ref{met_h}) on different copies of the spaces
${\bf S}^n$ and ${\bf H}^n$, corresponding to particles 1 and 2.

\subsection{Two-particle Hamiltonian on $\bf S^n\times\bf S^n$}\lb{S3Hamiltonian}

Given the point ${\bf x}_0\in F_r$ one can identify the
submanifold $F_r$ with the factor space $\SO(n+1)/\SO(n-1)$ by the
formula ${\bf x}=gK{\bf x}_0$, where $gK$ is the left coset of the
element $g$ in the group $\SO(n+1)$ w.r.t.\ its subgroup $K$. Let
$(r,y_1,\dots,y_{2n-1})$ be local coordinates in some neighborhood
$W$ of the point ${\bf x}_0\in Q_s$ and then
$(y_1,\dots,y_{2n-1})$ are the coordinates in the open subset
$W\bigcap F_r$ of the submanifold $F_r$. Then the metric
$\widetilde{g}_s$ in $W$ becomes\footnote{We suppose that this
metric does not contain summands proportional to $drdy_{i}$. It is
valid for the parametrization (\ref{AlphaBeta}) below. More
detailed consideration for a more general parametrization and an
arbitrary two-point homogeneous spaces $M$ can be found in
\cite{Shchep023}. The connection of the two-body Hamiltonian with
the algebra of invariant differential operators on the unit sphere
bundle over $M$ was described in \cite{Shch022}.}
$$\widetilde{g}_s=g_{rr}(r)dr^2+\sum\limits_{i,j=1}^{2n-1}g_{ij}(r,y_1,\dots,y_{2n-1})
dy_idy_j.$$ The second term in this formula is the restriction of
a metric $g_f$ from the submanifold $F_r$ onto the domain
$W\bigcap F_r$. Using the standard expression for the
Laplace-Beltrami operator trough local coordinates, one gets
\begin{equation}\lb{Hamcil}
\laplace_{\widetilde{g}_s}=\left(g_{rr}\det g_{ij}\right)^{-1/2}
\pd1{r}\left(\sqrt{g^{rr}\det
g_{ij}}\pd1{r}\right)+\laplace_{g_f}.
\end{equation}
To express the operator $\laplace_{g_f}$ on $F_r$ through the
generators of the Lie group $\SO(n+1)$ one can use the following
construction from \cite{Hel}.

Let $\Gamma$ be a Lie group and $\Gamma_0$ be its subgroup. The
group $\Gamma$ acts from the left on the homogeneous space
$\Gamma/\Gamma_0$. Left-invariant differential operators on the
space $\Gamma/\Gamma_0$ can be represented by left-invariant
differential operators on the group $\Gamma$ that are
simultaneously invariant w.r.t.\ the right action of the group
$\Gamma_0$. This representation is one to one up to operators
summands, vanishing while acting onto functions that are invariant
w.r.t.\ right $\Gamma_0$-shifts.

Indeed, functions on the factor space $\Gamma/\Gamma_0$ are in one
to one correspondence with functions on the group $\Gamma$ that
are invariant w.r.t.\ right $\Gamma_0$-shifts. This correspondence
is defined by the formula $\lambda:\;f\to \tilde f:=f\circ\pi$,
where $\pi:\;\Gamma\to \Gamma/\Gamma_0$ is the canonical
projection and $f$ is a function on the factor space
$\Gamma/\Gamma_0$. Let $D$ be a differential operator on $\Gamma$
that is invariant w.r.t.\ left $\Gamma$-shifts and right
$\Gamma_0$-shifts. Let $f$ be a smooth function on the factor
space $\Gamma/\Gamma_0$. Then the formula $\widetilde{D_u
f}=D\tilde f$ defines the correspondence $D\to D_u$, where $D_u$
is a differential operator on the space $\Gamma/\Gamma_0$,
invariant w.r.t.\ left $\Gamma$-shifts.

Let $e_1,\dots,e_{N}$ be a base of the Lie algebra of the group
$\Gamma,\quad N:=\dim\Gamma$ and let $L_{\gamma}$ and $R_{\gamma}$
denote respectively the left and right shifts by the element
$\gamma\in\Gamma$. The algebra of left invariant differential
operators on the group $\Gamma$ is generated over $\mathbb{R}$ by
left invariant vector fields $e^l_1,\dots,e_N^l$, where
$e_i^l(\gamma)=dL_{\gamma}(e_i),\;\gamma\in\Gamma,\;i=1,\dots,N$
\cite{Hel}.

Let now
$\Gamma=\SO(n+1),\;\Gamma_0=K\cong\SO(n-1),\;e_i^r(\gamma)=dR_{\gamma}(e_i),\;
i=1,\dots,N,\; N=(n+1)(n+2)/2,\;{\bf
x}_0=(r_1,\underbrace{0,\dots,0}_{n-1},r_2,
\underbrace{0,\dots,0}_{n-1})\in{\bf S}^n\times{\bf S}^n$, where
\begin{equation}\label{AlphaBeta}
r_1=\tan\left(\frac{m_2}{m_1+m_2}\arctan r\right), \quad
r_2=-\tan\left(\frac{m_1}{m_1+m_2}\arctan r\right).
\end{equation}

The set of Killing vector fields $X_{ij}^s,Y_i^s,\,i,j=1,\dots,n$
on the space ${\bf S}^n\times{\bf S}^n$, corresponding to
(\ref{basis_s}), coincides (up to a permutation) with the set
\begin{equation}\lb{corresp} \left\{\tilde e^r_i(\gamma x_0)=\left.\frac{d}{d\tau}
\right|_{\tau =0}\exp (\tau e_i)\gamma {\bf x}_0 \right\}^N_{i=1},
\quad {\bf x}_0={\bf x}_0(r),\quad 0<r<\infty,
\end{equation}
under the proper choice of the basis $e_1,\dots,e_N$. Let
$\laplace_f $ be a second order differential operator on the group
$\Gamma$ such that $\left(\laplace_f\right)_u=\laplace_{g_f}$.
Then it is left invariant and can be expressed in the
form\footnote{Here we consider vector fields as differential
operators of the first order.}
$$\left.\laplace_f\right|_{\gamma}=
\sum\limits_{i,j=1}^N\left.c^{ij}e_i^l\circ e_j^l\right|_{\gamma}+
\sum\limits_{i=1}^N\left.c^ie_i^l\right|_{\gamma},$$ where
$c^{ij}, c^i$ are constant on the submanifold $F_r$. Let $e$ be
the unit element of the group $\Gamma$. Obviously,
$\left.e_i^r\right|_{e}=\left.e_i^l\right|_{e},\quad i=1,\dots,N$
and
\begin{equation}\lb{delta_f}
\left.\laplace_f\right|_e= \sum\limits_{i,j=1}^N
\left.c^{ij}e_i^r\circ e_j^r\right|_{e}+ \sum\limits_{i=1}^N
\left.c^ie_i^r\right|_{e}.
\end{equation}
It yields
$$\left.\laplace_{g_f}\right|_{{\bf x}_0}=\sum\limits_{i,j=1}^N c^{ij}
\left.\tilde e_i^r\circ\tilde e_j^r\right|_{{\bf x}_0} +
\sum\limits_{i=1}^N\left.c^i\tilde e_i^r\right|_{{\bf x}_0}=:
\left.\laplace_{g_f}^{(2)}\right|_{{\bf x}_0}+
\left.\laplace_{g_f}^{(1)}\right|_{{\bf x}_0}.$$

One can find coefficients $c^{ij}$ in the following way. Consider
the ordered set of vectors $\left\{\left.Y_1^s\right|_{{\bf
x}_0},\dots,\left.Y_n^s\right|_{{\bf
x}_0},\left.X_{12}^s\right|_{{\bf x}_0},\dots,
\left.X_{1n}^s\right|_{{\bf x}_0}\right\}$ as a base in the linear
space $T_{{\bf x}_0}F_r$. Let $\left\{
Y^1,\dots,Y^n,X^2,\dots,X^n\right\}$ be the dual basis. Then
\begin{align*}
\left.g_f\right|_{{\bf x}_0}&=\sum\limits_{i=2}^n\left[Y^1
\otimes\left(\alpha_iY^i+\beta_iX^i\right)+\sum\limits_{j=2}^n\left(\alpha_{ij}
Y^i\otimes Y^j+\beta_{ij}X^i\otimes X^j+\gamma_{ij}Y^i\otimes
X^j\right)\right]\\&+aY^1\otimes Y^1,
\end{align*}
where
\begin{align*}
a&=\left.\widetilde{g}\left(Y^s_1,Y^s_1\right)\right|_{{\bf
x}_0}=2R^2(m_1+m_2), \\
\alpha_i&=\left.\widetilde{g}\left(Y^s_1,Y^s_i\right)\right|_{{\bf
x}_0}=0, \\ \beta_i
&=\left.\widetilde{g}\left(Y^s_1,X^s_{1i}\right)\right|_{{\bf
x}_0}=0,\quad i=2,\dots,n,
\end{align*}
\begin{align}\lb{coeff_1}\alpha_{ij} &=
\left.\widetilde{g}\left(Y^s_i,Y^s_j\right)\right|_{{\bf
x}_0}=2R^2\sum\limits_{k=1}^2\frac{m_k\left(1-r_k^2
\right)^2}{\left(1+r_k^2\right)^2}\delta_{ij}, \nonumber \\
\beta_{ij}&=\left.\widetilde{g}\left(X^s_{1i},X^s_{1j}
\right)\right|_{{\bf x}_0} = 8R^2\sum\limits_{k=1}^2\frac{m_k
r_k^2}{\left(1+r_k^2\right)^2} \delta_{ij}, \\
\gamma_{ij}&=\left.\widetilde{g}\left(Y^s_i,X^s_{1j}\right)\right|_{{\bf
x}_0}= 4R^2\sum\limits_{k=1}^2\frac{m_k r_k(1-r_k^2)}
{\left(1+r_k^2\right)^2}\delta_{ij},\quad i,j=2,\dots,n. \nonumber
\end{align}
Therefore one gets
\begin{align}\lb{oper_x_0}
\left.\laplace_{g_f}^{(2)}\right|_{{\bf x}_0}=\frac1a
\left.\left(Y_1^s\right)^2\right|_{{\bf
x}_0}+\frac12\sum\limits_{k=2}^n\left[A_s \left(X_{1k}^s\right)^2
+ C_s\left(Y_k^s\right)^2-B_s\left\{X_{1k}^s,
Y_k^s\right\}\right]_{{\bf x}_0},
\end{align}
where $\left\{\cdot,\cdot\right\}$ denotes the anticommutator and
functions $A_s,B_s,C_s$ have the form
\begin{align*}
A_s &=\frac{m_1(1-r_1^2)^2(1+r_2^2)^2+m_2(1+r_1^2)^2(1-r_2^2)^2}
{4R^2m_1m_2(r_1-r_2)^2(1+r_1r_2)^2}, \\ B_s
&=\frac{m_1r_1(1-r_1^2)(1+r_2^2)^2+m_2r_2(1-r_2^2)(1+r_1^2)^2}
{2R^2m_1m_2(r_1-r_2)^2(1+r_1r_2)^2}, \\ C_s
&=\frac{m_1r_1^2(1+r_2^2)^2+m_2r_2^2(1+r_1^2)^2}
{R^2m_1m_2(r_1-r_2)^2(1+r_1r_2)^2}.
\end{align*}
These functions can be expressed also through the coordinate $r$:
\begin{align*}
A_s(r) &=\frac1{R^2}\left(\frac{(1 + r^2)^2}{8mr^2} +
\frac{1-r^4}{8mr^2} \cos\zeta + \frac{1 + r^2}{4m_1m_2r}(m_1-m_2)
\sin\zeta\right), \\ B_s(r)
&=-\frac1{4R^2}\left(\frac{m_2-m_1}{m_1m_2r}(1 + r^2)\cos\zeta +
\frac{1-r^4}{2mr^2}\sin\zeta \right), \\ C_s(r)
&=\frac1{R^2}\left(\frac{(1 + r^2) ^2}{8mr^2}-\frac{1-r^4}{8mr^2}
\cos\zeta - \frac{1 + r^2}{4m_1m_2r}(m_1-m_2)\sin\zeta\right),\\
\zeta&:=2\frac{m_1-m_2}{m_1 + m_2} \arctan r, \quad
m:=\frac{m_1m_2}{m_1 + m_2}.\end{align*}

The operators $\left.\laplace_{g_f}\right|_{{\bf x}_0}$ and
$\left.\laplace_{g_f}^{(2)}\right|_{{\bf x}_0}$ (see
(\ref{oper_x_0})) are invariant w.r.t.\ reflections $T_k:\; x_k\to
-x_k, x_j\to x_j, j\ne k,\,j=1,\ldots,n,\,k=2,\ldots,n$ of the
sphere ${\bf S}^n$. Therefore the operator
$\left.\laplace_{g_f}^{(1)}\right|_{{\bf x}_0}$ is also invariant
w.r.t.\ these reflections. Reflections $T_{k}$ changes signs of
the vector fields $\left.X_{1k}^{s}\right|_{{\bf
x}_0},\left.Y_{k}^{s}\right|_{{\bf x}_0},\,k=2,\ldots,n$,
therefore the operator $\left.\laplace_{g_f}^{(1)}\right|_{{\bf
x}_0}$ is proportional to $\left.Y_{1}^{s}\right|_{{\bf x}_0}$.
The more accurate analysis (see \cite{Shchep023}) shows that the
operator $\laplace_{g_f}^{(1)}$ vanishes.

If we denote by $Y_1^{s,l}, X_k^{s,l}, Y_k^{s,l} $ left invariant
vector fields on the group $\SO(n+1)$, corresponding to vectors
$\left.Y_1^s\right|_{{\bf x}_0},\left.X_{1k}^s\right|_{{\bf
x}_0},\left.Y_k^s\right|_{{\bf x}_0},\,k=2,\dots,n$, we get
\begin{align}\lb{delta}
\laplace_f=\frac1a D_{0}^{2}+\frac12A_sD_{2}+
\frac12C_sD_{1}+B_sD_{3},
\end{align}
where operators $D_{0},D_{1},D_{2}$ and $D_{3}$ have the form
$$
D_0=Y_1^{s,l},\;D_1=\sum\limits_{k=2}^n\left(Y_k^{s,l}\right)^2,\;
D_2=\sum\limits_{k=2}^n\left(X_{k}^{s,l}
\right)^2,\;D_3=-\frac12\sum\limits_{k=2}^n\left\{X_{k}^{s,l},Y_k^{s,l}\right\}.
$$

By direct calculations in the universal enveloping algebra
$U(\mathfrak{so}(n+1))$, one can get (see \cite{Shch022}) the
following commutator relations for the operators
$D_{0},\ldots,D_{3}$
\begin{align}\label{CommRelations}
&[D_{0},D_{1}]=-2D_{3},\,[D_{0},D_{2}]=2D_{3},\,[D_{0},D_{3}]=D_{1}-D_{2},\,
[D_{1},D_{2}]=-2\{D_{0},D_{3}\},\\ &[D_{1},D_{3}]=-\{D_{0},D_{1}\}
+\frac{(n-1)(n-3)}2D_{0},\,[D_{2},D_{3}]=\{D_{0},D_{2}\}-\frac{(n-1)(n-3)}2D_{0}.\notag
\end{align}

Thus we found the operator $\laplace_f$ up to summands, annulled
by functions that are invariant w.r.t.\ right $\Gamma_0$-shifts.

We are to find now the first term in expression (\ref{Hamcil}). At
the point ${\bf x}_0$ one has
$$\pd1{r}=\frac{m_2}{m_1+m_2}\frac{1+r_1^2}{1+r^2}\pd1{r_1}-
\frac{m_1}{m_1+m_2}\frac{1+r_2^2}{1+r^2}\pd1{r_2}$$ and therefore
\begin{equation}\label{grr}
g_{rr}=\widetilde{g}\left(\pd1 r,\pd1 r\right)=
\frac{8R^2m_1m_2}{(m_1+m_2)(1+r^2)^2}.
\end{equation}

Due to formulas (\ref{coeff_1}) one gets
$$\laplace_{\widetilde{g}_{s}}=\frac{(1+r^2)^n}{8mR^2r^{n-1}}\pd1{r}\left(
\frac{r^{n-1}}{(1+r^2)^{n-2}}\pd1 r\right)+\laplace_{g_f},$$ where
the first term is the radial part of the one particle Hamiltonian
with the mass $m$.

Direct calculation at the point ${\bf x}_0$ gives for the measure
$d\mu_s$, corresponding to the metric $\widetilde{g}_{s}$ in the
space $Q_s$, the following expression
$$\left.d\mu_s\right|_{{\bf x}_0}=\frac{r^{n-1}}{(1+r^2)^n}dr\land Y^1\land\dots
\land Y^n\land X^2\land\dots\land X^n
$$
up to a constant factor.

The measure $d\mu_s$ is left invariant w.r.t.\ the group
$\SO(n+1)$ and therefore it can be represented in the form
$d\mu_s=d\nu_s\otimes d\mu_f $, where $d\nu_s=r^{n-1}dr/(1+r^2)^n$
is a measure on $\mathbb{R}_+=(0,\infty)$, the same as for the one
particle case, and $d\mu_f$ is a measure on $\SO(n+1)/K$ left
invariant w.r.t.\ the group $\SO(n+1)$.

Each Lie group admits unique (up to a constant factor)
left-invariant and right-invariant measures (Haar measures
\cite{Kir}). For the groups under consideration $\SO(n+1)$ and
$\Or_{0}(1,n)$ such measures are two-side invariant. Hence there
exist a unique two-side invariant measure $d\eta_s$ on the group
$\SO(n+1)$ such that the integral of any integrable function $f$
on the space $\SO(n+1)/K$ w.r.t.\ the measure $d\mu_f$ equals the
integral of the function $\tilde f$ on the group $\SO(n+1)$
w.r.t.\ the measure $d\eta_s$.

\begin{Def}\label{def1}
For a subgroup $\Gamma_0$ of a Lie group $\Gamma$ denote by
$\mathcal{L}^2(\Gamma,\Gamma_0,d\eta)$ the space of
square-integrable functions on the group $\Gamma$ (w.r.t.\ the
measure $d\eta$ on $\Gamma$) that are right invariant w.r.t.\
$\Gamma_0$-shifts.
\end{Def}

\begin{theore}\label{th1}
The free two-particle Hamiltonian on the sphere ${\bf S}^n$ can be
considered as the self-adjoint differential operator (on the
manifold $\widetilde{Q}_s=\mathbb{R}_+\times\SO(n+1)$)
\begin{equation}\lb{H_0^s}
\widehat{H}_0^s=-\frac{(1+r^2)^n}{8mR^2r^{n-1}}\pd1{r}\left(
\frac{r^{n-1}}{(1+r^2)^{n-2}}\pd1 r\right)-\laplace_f,
\end{equation}
with the domain
$$ D_s:=D^{(1)}_s\otimes D^{(2)}_s\subset\mathcal{H}_s:=\mathcal{L}^2
\left(\mathbb{R}_+,d\nu_s\right)\otimes\mathcal{L}^2\left(\SO(n+1),K,d\eta_s\right),$$
where
\begin{align*}
D^{(1)}_s&:=\left\{\phi\in\mathcal{L}^2\left(\mathbb{R}_+,d\nu_s\right)|
\laplace_s^{(1)}\phi\in\mathcal{L}^2\left(\mathbb{R}_+,d\nu_s\right)\right\},\\
D^{(2)}_s&:=\left\{\phi\in\mathcal{L}^2\left(\SO(n+1),K,d\eta_s\right)|
\laplace_f\phi\in\mathcal{L}^2\left(\SO(n+1),K,d\eta_s\right)\right\},\\
\laplace^{(1)}_s&:=-\frac{(1+r^2)^n}{r^{n-1}}\pd1{r}\left(\frac{r^{n-1}}
{(1+r^2)^{n-2}}\pd1 r\right),
\end{align*}
the subgroup $K$ is isomorphic to the group $\SO(n-1)$, and
$d\eta_s$ is a unique (up to a constant factor)  two-side
invariant measure on the group $\SO(n+1)$. It means that there
exists an isometry of the initial space of functions
$\mathcal{L}^2\left(Q_s,d\mu_s\right)$ onto the space
$\mathcal{H}_s$ that generates the isomorphism of Hamiltonians.
The space $D_s$ is everywhere dense in $\mathcal{H}_s$.
\end{theore}
\begin{proof}
Expression (\ref{Hamcil}) represents the Hamiltonian
$\widehat{H}_0^s$ via coordinates, corresponding with the
representation of the space $Q_s$ as the direct product
$\mathbb{R}_+\times\SO(n+1)/\SO(n-1)$ up to the zero measure set
$F^s_0\cup F^s_{\infty}$, which is inessential when studying
functions that are integrable over this measure. Therefore
$$\mathcal{L}^2\left(Q_s,d\mu_s
\right)=\mathcal{L}^2\left(\mathbb{R}_+,d\nu_s\right)\otimes\mathcal{L}^2\left(
\SO(n+1)/\SO(n-1), d\mu_f\right).
$$
The isometry $\lambda: f\to\tilde f$ of spaces
$$\mathcal{L}^2\left(\SO(n+1)/\SO(n-1),
d\mu_f\right)\,\text{and}\;\mathcal{L}^2\left(\SO(n+1),\SO(n-1),
d\eta_s\right)$$ generates the isometry $\id\otimes\lambda$ of
spaces
$\mathcal{L}^2\left(\mathbb{R}_+,d\nu_s\right)\otimes\mathcal{L}^2\left(\SO(n+1)/\SO(n-1),
d\mu_f\right)$ and $\mathcal{H}_s$. Calculations above imply that
the isometry $\id\otimes\lambda$ transforms operator
(\ref{Hamcil}) into operator (\ref{H_0^s}) and the space $W_s$
into the space $D_s$.
\end{proof}

\begin{Rem}\label{rem1}
In the case $n=2$ this result can be obtained by treating a basis
in the Lie algebra $\mathfrak{so}(3)$ as a moving frame on the
submanifold $F_r$ \cite{Shch2}. For $n>2$ this is impossible since
the $\SO(n+1)$-action on $F_r$ is not free and the projection of
left-invariant vector fields from $\SO(n+1)$ onto
$\SO(n+1)/\SO(n-1)$ is not well defined. By lifting the
Hamiltonian onto the symmetry group one can express it through
group generators.
\end{Rem}

\subsection{Two-particle Hamiltonian on the space
$\bf H^n\times\bf H^n$}\lb{H3Hamiltonian}

The formal substitution $x_j\to -ix_j, r\to -ir, R\to iR,
j=1,\dots,n,$ (here $i$ is the imaginary unit) transforms objects
on the sphere ${\bf S}^n$ into objects on the hyperbolic space
${\bf H}^n$ (see also \cite{Shch1}). Therefore from results of the
previous section one can obtain the two-particle Hamiltonian on
the space ${\bf H}^n\times\bf H^n$:
\begin{align}\lb{H_o^h}
\widehat{H}_0^h=-\frac{(1-r^2)^n}{8mR^2r^{n-1}}\pd1{r}\left(
\frac{r^{n-1}}{(1-r^2)^{n-2}}\pd1 r\right)- \frac1a\D_0^2 -
\frac12A_h\D_2-\frac12C_h\D_1-B_h\D_3,
\end{align}
where $0<r<1$,
$$\D_0=Y_1^{h,l},\;D_1=\sum\limits_{k=2}^n\left(Y_k^{s,l}\right)^2,\;
\D_2=\sum\limits_{k=2}^n\left(X_{k}^{h,l}
\right)^2,\;\D_3=-\frac12\sum\limits_{k=2}^n\left\{X_{k}^{h,l},Y_k^{h,l}\right\},
$$
vector fields $X_k^{h,l}, Y_k^{h,l}$ relate to vector fields
(\ref{basis_h}) in the same way as vector fields $X_k^{s,l},
Y_k^{s,l}$ relate to vector fields (\ref{basis_s}), and
\begin{align*}
A_h(r) &=\frac1{R^2}\left(\frac{(1 - r^2)^2}{8mr^2} +
\frac{1-r^4}{8mr^2}\cosh\zeta - \frac{1-r^2}{4m_1m_2r}(m_1-m_2)
\sinh\zeta\right), \\
B_h(r)&=-\frac1{4R^2}\left(\frac{(m_2-m_1)}{m_1m_2r}(1 -
r^2)\cosh\zeta + \frac{1-r^4}{2mr^2}\sinh\zeta \right), \\
C_h(r)&=-\frac1{R^2}\left(-\frac{(1 - r^2)
^2}{8mr^2}+\frac{1-r^4}{8mr^2} \cosh\zeta - \frac{1 -
r^2}{4m_1m_2r}(m_1-m_2)\sinh\zeta\right),
\\ \zeta&:=2\frac{m_1-m_2}{m_1 + m_2}\arctanh r.
\end{align*}

Commutator relations for operators $\D_{0},\D_{1},\D_{2},\D_{3}$
have the form (see \cite{Shch022})
\begin{align}\lb{ComRelationH}
&[\D_{0},\D_{1}]=2\D_{3},\,[\D_{0},\D_{2}]=2\D_{3},\,[\D_{0},\D_{3}]=\D_{1}+\D_{2},\,
[\D_{1},\D_{2}]=-2\{\D_{0},\D_{3}\},\\
&[\D_{1},\D_{3}]=-\{\D_{0},\D_{1}\}-\frac{(n-1)(n-3)}2\D_{0},\,
[\D_{2},\D_{3}]=\{\D_{0},\D_{2}\}-\frac{(n-1)(n-3)}2\D_{0}.\notag
\end{align}

\begin{theore}\label{th2}
The free two-particle Hamiltonian on the space ${\bf H}^n$ can be
considered as the self-adjoint differential operator (\ref{H_o^h})
on the manifold $\widetilde{Q}_h=I\times\Or_{0}(1,n)$ with the
domain
$$D_h:=D^{(1)}_h\otimes D^{(2)}_h\subset\mathcal{H}_h:=\mathcal{L}^2
\left(\mathbb{R}_+,d\nu_h\right)\otimes\mathcal{L}^2\left(\Or_{0}(1,n),K,d\eta_h\right),$$
where $K=\SO(n-1)$,
\begin{align*}
D^{(1)}_h&:=\left\{\phi\in\mathcal{L}^2\left(\mathbb{R}_+,d\nu_h\right)|
\laplace_h^{(1)}\phi\in\mathcal{L}^2\left(\mathbb{R}_+,d\nu_h\right)\right\},\\
D^{(2)}_h&:=\left\{\phi\in\mathcal{L}^2\left(\Or_{0}(1,n),K,d\eta_h\right)|
\laplace_h\phi\in\mathcal{L}^2\left(\Or_{0}(1,n),K,d\eta_h\right)\right\},\\
\laplace^{(1)}_h&:=-\frac{(1-r^2)^n}{r^{n-1}}\pd1{r}\left(\frac{r^{n-1}}
{(1-r^2)^{n-2}}\pd1 r\right), \;
d\nu_h=\frac{r^{n-1}dr}{(1-r^2)^n}, \\ \laplace_h&:=-\frac1a\D_0^2
- \frac12A_h\D_2-\frac12C_h\D_1 - B_h\D_3,
\end{align*}
and $d\eta_h$ is a unique (up to a constant factor) two-side
invariant measure on the group $\Or_{0}(1,n)$.
\end{theore}

The proof is analogous to the proof of theorem \ref{th1}.

\section{Self-adjointness of two-particle Hamiltonians}

In Euclidean space the self-adjointness of many-particle
Hamiltonians with pairwise interacting particles is usually proved
using the Galilei invariance, which has no analog in the spaces
${\bf S}^n$ and ${\bf H}^n$ \cite{Sim_R}. The self-adjointness of
one-particle Hamiltonians with singular potential unbounded from
below can be proved using the perturbation theory for the
corresponding quadratic forms. The key point of the proof is the
following estimate, called the "uncertainty principle" in
\cite{Sim_R}:
$$(U\psi,\psi)\leqslant\|\nabla\psi\|^2,$$
where $(\cdot,\cdot)$ is the scalar product in the space
$\mathcal{L}^2(\mathbb{R}^3)$, $\nabla$ is the gradient operator,
and $U$ is a potential. To derive this estimate for spaces ${\bf
S}^n$ and ${\bf H}^n$ one needs some modification of the proof
w.r.t.\ the Euclidean case. Instead of tending to full generality,
we are mostly interested in physically significant potentials.

From the self-adjointness of the free two-particle Hamiltonian
with domain (\ref{DomSelfAdjointness}) we shall prove the
self-adjointness of the two-particle Hamiltonian with an
interaction using the perturbation theory for the quadratic forms.

Let $(\cdot,\cdot)$ be the scalar product on fibers of the
cotangent bundle $T^{*}Q_s$, generated by the metric $\tilde g_s$,
$\|\cdot\|$ and $\nabla$ be respectively the corresponding norm
and the gradient operator. Let also $f,\psi\in C^{\infty}(Q_s)$ be
real functions\footnote{All functional spaces are assumed to
comprise complex-valued functions.} such that $f$ is constant on
submanifolds $F^s_r$, i.e.\ $f=f(r)$. Then it holds
\begin{align*}
\left\|\nabla\psi\right\|^2 &=\left\|\frac{\nabla(f\psi)}f-
\frac{\psi\nabla f}f\right\|^2 \geqslant\frac{\psi^2\|\nabla
f\|^2}{f^2}- \frac{2\psi\left(\nabla f, \nabla(f\psi)\right)}{f^2}
\\ &=\frac{\psi^2\tilde
g^{rr}_s}{f^2}(f')^2-\frac{2\psi}{f^2}\tilde g^{rr}_sf'
\pd1{r}\left(f\psi\right)
\end{align*}
Equation (\ref{grr}) implies that $\tilde
g^{rr}=(1+r^2)^2/(8R^2m)$. Integrating over $Q_s$ with the measure
$d\mu_s$, one gets
\begin{align}\begin{split}
&\int\limits_{Q_s}\|\nabla\psi\|^2d\mu_s \geqslant
\frac1{8R^2m}\int\limits_{Q_s}\left[\frac{(f')^2\psi^2}{f^2}-\frac{2\psi
f'} {f^2}\pd1{r}\left(f\psi\right)\right](1+r^2)^2d\mu_s
 \\ &=\frac1{8R^2m}\int\limits_{Q_s}\frac{(f')^2\psi^2}{f^2}(1 +
r^2)^2d\mu_s-\frac1{4R^2m}\int\limits_{F^s_r}\int\limits_0^{\infty}\frac{\psi
f'r^{n-1}} {f^2(1+r^2)^{n-2}}\pd1{r}\left(f\psi\right)drd\mu_f.
\end{split}\lb{ineq1}
\end{align}
We now want to find a function $f$ such that for every smooth
function $\psi$, the function
$$\frac{\psi f'r^{n-1}}{f^2(1+r^2)^{n-2}}\pd1{r}\left(f\psi\right)$$
has the form $\pd1{r}\left(\phi(r)\psi^2\right)$. Then the last
integral in (\ref{ineq1}) can vanish identically. Solving the
system of equations
$$\frac{(f')^2r^{n-1}}{f^2(1+r^2)^{n-2}}=\phi',\quad
\frac{f'r^{n-1}}{f(1+r^2)^{n-2}}=2\phi,$$ one gets
\begin{align}\label{asyptotics}
\phi(r)&=-\left[4\int\frac{(1+r^2)^{n-2}}{r^{n-1}}dr\right]^{-1},\quad
\phi(r)\sim\left\{\begin{array}{c}(n-2)r^{n-2}/4,\;n\geqslant 3 \\
\left(4|\ln r|\right)^{-1},\; n=2 \\ \end{array}\right.,\; r\to
0,\notag
\\ \phi(r)&\sim\left\{\begin{array}{c}-(n-2)r^{2-n}/4,\;n\geqslant 3 \\
-\left(4|\ln r|\right)^{-1},\; n=2 \\
\end{array}\right.,\; r\to \infty,\\
\left(\frac{f'}f\right)^2(r)&\sim\left\{\begin{array}{c}(n-2)^2/
(4r^2),\; n\geqslant 3 \\ \left(4r^2\ln^2r\right)^{-1},\; n=2 \\
\end{array} \right.,\; r\to 0,\; r\to \infty.\notag
\end{align}

It is easy to verify that for any choice of the integration
constant the function $\phi(r)$ is discontinues at some point,
which does not occur in the Euclidean case for $n\geqslant 3$. Let
$\omega_{\delta}:=\left\{x\in Q_s| r(x)< \delta\right\}$ and
$\omega'_{\delta}:=\left\{x\in Q_s| r(x)>\delta^{-1} \right\}$.
Choose $\delta>0$ in such a way that the function $\phi(r)$ is
continues in the domain
$\left(\omega_{2\delta}\cup\omega'_{2\delta}\right)\setminus\left(F^s_0\cup
F^s_{\infty}\right)$. On the space $Q_s\setminus\left(F^s_0\cup
F^s_{\infty} \right)$ the continuity domains of the functions
$f'/f$ and $\phi(r)$ coincide; moreover both functions are nonzero
for $r\ne 0,\infty$. We set
$$ u_n(r)=\left\{\begin{array}{c}r^{-2}+r^2,\;n\geqslant 3 \\
\left(r^{-2}+r^2\right)/\ln^2r,\; n=2 \\ \end{array}\right.,$$ and
choose the constant $\kappa>0$ such that the inequality
$$\kappa u_n(r)\leqslant\frac{(1+r^2)^2\left(f'\right)^2}{8R^2mf^2}$$
holds. Then due to (\ref{ineq1}) and (\ref{asyptotics}) one gets
the following inequality
\begin{equation}\lb{ineq_main}
\kappa\int\limits_{Q_s}u_n(r)|\psi|^2d\mu_s\leqslant
\int\limits_{Q_s}\|\nabla\psi\|^2d\mu_s
\end{equation}
for the function $\psi\in C^{\infty}\left(Q_s\right)$ with
$\supp\,\psi\subset\omega_{\delta}\cup\omega'_{\delta}$. Writing
inequality (\ref{ineq_main}) separately for the real and imaginary
parts of a complex-valued function, we obtain (\ref{ineq_main})
for an arbitrary function $\psi\in C^{\infty}\left(Q_s\right)$
with $\supp\psi\subset\omega_{\delta}\cup\omega'_{\delta}$.

\begin{theore}\label{th3}
Let $U$ be a real function that is smooth in the domain
$Q_s\setminus\left(F^s_0\cup F^s_{\infty}\right)$ and satisfies
the estimate $U=o\left(u_n(r)\right)$ as $r\to 0,\infty$,
uniformly w.r.t.\ coordinates on the second factor of the direct
product $\mathbb{R}_+\times\SO(n+1)/\SO(n-1)$, representing $Q_s$
up to the zero measure set $F^s_0\cup F^s_{\infty}$. Then the
two-particle Hamiltonian $\widehat{H}_s$ is essentially
self-adjoint in any domain of essential self-adjointness of the
free Hamiltonian $\widehat{H}^s_0$. In particular, $\widehat{H}_s$
is essentially self-adjoint in the domain
$C^{\infty}\left(Q_s\right)$.
\end{theore}
\begin{proof}
Theorem X.17 from \cite{Sim_R} states that it is suffice to prove
that the inequality
$$\int\limits_{Q_s}|U||\psi|^2d\mu_s\leqslant a\int\limits_{Q_s}\bar\psi\widehat{H}^s_0
\psi d\mu_s +b\int\limits_{Q_s}|\psi|^2d\mu_s,
$$
is valid for all $\psi\in C^{\infty}(Q_s)$, where
$0<a<1,\;b\in\mathbb{R}$. Fixing a constant $a\in (0,1)$, we
choose the function $\chi\in C^{\infty}(Q_s)$ such that
$\supp\,\chi\subset \omega_{\delta}\cup\omega'_{\delta},\;
\chi|_{\omega_{\delta/2}\cup\omega'_{\delta/2}}\equiv 1,\;
0\leqslant\chi\leqslant 1$. Now let
$0<\varepsilon\leqslant\delta/2$ such that $|U(x)|\leqslant
a\kappa
u_n(r(x))/2,\;x\in\omega_{\varepsilon}\cup\omega'_{\varepsilon}$.
Let also
$$c:=\sup\limits_{x\in
Q_s\setminus(\omega_{\varepsilon}\cup\omega'_
{\varepsilon})}^{}\,|U(x)|\;\text{and}\;\psi\in C^{\infty}(Q_s).$$
Then
$$\int\limits_{Q_s}|U||\psi|^2d\mu_s\leqslant
\frac
a2\kappa\int\limits_{\omega_{\varepsilon}\cup\omega'_{\varepsilon}}
u_n(x)|\chi\psi|^2 d\mu_s +c\int\limits_{Q_s}|\psi|^2d\mu_s.
$$
Applying estimate (\ref{ineq_main}) to the first integral, one
gets
$$\kappa\int\limits_{\omega_{\varepsilon}\cup\omega'_{\varepsilon}}
u_n(x)|\chi\psi|^2 d\mu_s \leqslant
\int\limits_{Q_s}\left\|\nabla(\chi\psi)\right\|^2 d\mu_s
\leqslant 2\int\limits_{Q_s}\left(|\nabla\psi|^2 +
|\nabla\chi|^2|\psi|^2\right) d\mu_s.$$ Hence
$$\int\limits_{Q_s}|U||\psi|^2d\mu_s\leqslant
a\int\limits_{Q_s}\left\|\nabla\psi\right\|^2 d\mu_s +
b\int\limits_{Q_s}|\psi|^2d\mu_s =
a\int\limits_{Q_s}\bar\psi\widehat{H}^s_0 \psi d\mu_s
+b\int\limits_{Q_s}|\psi|^2d\mu_s,$$ where
$b=c+2\sup\limits_{Q_s}\left(|\nabla\chi|^2\right)\;\square$.
\end{proof}

An analogous result is valid for the space ${\bf H}^n$.

\begin{theore}
Let $U$ be a real function that is smooth in the domain
$Q_h\setminus F^h_0$ and satisfies the estimate
$U=o\left(u_n(r)\right)$ as $r\to 0,\infty$, uniformly w.r.t.\
coordinates on the second factor of the direct product
$I\times\Or_{0}(1,n)/\SO(n-1)$, representing $Q_h$ up to the zero
measure set $F^h_0$. Then the two-particle Hamiltonian
$\widehat{H}_h$ is essentially self-adjoint in any domain of
essential self-adjointness of the free Hamiltonian
$\widehat{H}^h_0$. In particular, $\widehat{H}_h$ is essentially
self-adjoint in the domain $C^{\infty}\left(Q_h\right)$.
\end{theore}

\section{The spectrum of the operator $\widehat{H}_s$}

It is known (see, for instance, \cite{Vil}) that all (enumerable
set) irreducible representations of a compact Lie group $\Gamma$
are finite dimensional and are contained in its regular
representation by left or right shifts in the space
$\mathcal{L}^2\left(\Gamma,d\eta\right)$, where $\eta$ is a
two-side invariant measure on the group $\Gamma$. Let
\begin{equation}\label{l2DirectExpansion}
\mathcal{L}^2\left(\SO(n+1),d\eta_{s}\right)=\bigoplus_{k}\mathcal{T}_{k}
\end{equation}
be the decomposition of the right regular representation of the
group $\SO(n+1)$ into irreducible ones. Restricting decomposition
(\ref{l2DirectExpansion}) onto the subspace
$\mathcal{L}^2\left(\SO(n+1),K,d\eta_{s}\right)$ of the space
$\mathcal{L}^2\left(\SO(n+1),d\eta_{s}\right)$ one gets
\begin{equation}\label{l2DirectExpansionK}
\mathcal{L}^2\left(\SO(n+1),K,d\eta_{s}\right)=\bigoplus_{k}\mathcal{T}_{k}',
\end{equation}
where subspaces $\mathcal{T}_{k}'\subset\mathcal{T}_{k}$ consist
of functions annulled by left invariant vector fields on
$\SO(n+1)$ generated by elements from the Lie algebra
$\mathfrak{k}$ of the group $K\cong\SO(n-1)$. Differential
operators in the space
$\mathcal{L}^2\left(\SO(n+1),K,d\eta_{s}\right)$, invariant
w.r.t.\ left $\SO(n+1)$-shifts, conserve subspaces
$\mathcal{T}_{k}'$.

Explicit (and rather cumbersome) expressions for the action of
infinitesimal generators of the group $\SO(n)$ on basis elements
(described by the Gelfand-Tsetlin schemes) of its irreducible
representations were found in \cite{Ott1}, \cite{Ott2}.

Thus the matrix ordinary differential operator that corresponds to
the restriction of $\widehat{H}_s$ onto the subspace
$\mathcal{L}^2\left(\mathbb{R}_+,d\nu_s\right)\otimes\mathcal{T}_k'$
can be found explicitly. It is interesting to find one-dimensional
subspaces $\widetilde{\mathcal{T}}_k\subset\mathcal{T}_k'$ for
which the spaces
$\mathcal{L}^2\left(\mathbb{R}_+,d\nu_s\right)\otimes\widetilde{\mathcal{T}}_k$
are invariant w.r.t.\ the operator $\widehat{H}_s$. In this case
one can found separate spectral ordinary differential equations
for the two-particle Hamiltonian on the space ${\bf S}^n$ that can
be solved explicitly for some potentials $U(r)$.

In the case $m_1\ne m_2$ formulas (\ref{delta}) and (\ref{H_0^s})
imply that for the realization of this program it is sufficient to
find common eigenfunctions for operators $D_0^{2},D_1,D_2,D_3$ in
the space $\mathcal{T}_k$. In the case $m_1=m_2$ one has
$B_s(r)\equiv 0$ and it is sufficient to find common
eigenfunctions for operators $D_0^{2},D_1,D_2$.

Due to the lack of a general methods for finding common
eigenvectors of noncommutative operators, we restrict ourselves
with the case of the three dimensional sphere ${\bf S}^3$. Recall
that the two dimensional sphere ${\bf S}^2$ was considered from
this point of view in \cite{Shch2}.

The base $L_1,L_2,L_3,G_1,G_2,G_3$ of the Lie algebra
$\mathfrak{so}(4)$ defined as
\begin{align}\begin{split}
L_1&=\frac12(X^s_{23}+Y^s_1),\;L_2=\frac12(X^s_{31}+Y^s_2),\;
L_3=\frac12(X^s_{12}+Y^s_3),\\ G_1&=\frac12(X^s_{23}-Y^s_1),\;
G_2=\frac12(X^s_{31}-Y^s_2),\;G_3=\frac12(X^s_{12}-Y^s_3)
\end{split}\label{NewBase}\end{align}
corresponds to the decomposition
$\mathfrak{so}(4)=\mathfrak{su}(2)\oplus\mathfrak{su}(2)$. The
corresponding commutator relations are\footnote{Commutative
relations for elements of a Lie algebra and commutative relations
of corresponding Killing vector fields (considered as differential
operators of the first order) differ by a sign for a left action
of an isometry group.}
$$ [L_l, L_j] =\sum_{k=1} ^3\varepsilon_{ljk} L_k,\;
   [G_l, G_j] =\sum_{k=1} ^3\varepsilon_{ljk} G_k,\;
   [L_l, G_j] =0,\; l,j=1,2,3,$$
where $\varepsilon_{ljk}$ is a totally antisymmetric tensor such
that $\varepsilon_{123}=1$. Let
\begin{equation*}
T_{\pm}=iL_{2}\pm L_{3},\,T_{0}=-iL_{1},\,W_{\pm}=iG_{2}\pm
G_{3},\,W_{0}=-iG_{1}
\end{equation*}
be a base in the complexification of the Lie algebra
$\mathfrak{so}(4)$, where $i$ is the imaginary unit. Then one has
the following commutative relations
\begin{gather*}
[T_{0},T_{+}]=T_{+},\,[T_{0},T_{-}]=-T_{-},\,[T_{+},T_{-}]=2T_{0},\\
[W_{0},W_{+}]=W_{+},\,[W_{0},W_{-}]=-W_{-},\,[W_{+},W_{-}]=2W_{0}.
\end{gather*}
Evidently, operators from different triples commute with each
other.

Since the group $\SU(2)\times\SU(2)$ is the double covering of the
group $\SO(4)$ the representation theory for the latter one can be
derived from the representation theory for the group $\SU(2)$.

Let $U_{\ell_{1}},\ell_{1}=0,\frac12,1,\frac32,\ldots$ be the
unitary space of irreducible representation
$\mathcal{T}_{\ell_{1}}$ for the group $\SU(2)$, generated by
elements $L_{1},L_{2},L_{3}$. This space has a base
$\psi^{\ell_{1}}_{n_{1}},\,n_{1}=-\ell_{1},-\ell_{1}+1,\ldots,\ell_{1}+1,\ell_{1}$,
satisfying relations (\cite{Vil})
\begin{align*}
T_{0}\psi_{n_{1}}^{\ell_{1}}&=n_{1}\psi_{n_{1}}^{\ell_{1}},\;
T_{+}\psi_{n_{1}}^{\ell_{1}}=-\sqrt{(\ell_{1}-n_{1})(\ell_{1}+n_{1}+1)}\psi_{n_{1}+1}^{\ell_{1}},\\
T_{-}\psi_{n_{1}}^{\ell_{1}}&=-\sqrt{(\ell_{1}+n_{1})(\ell_{1}-n_{1}+1)}\psi_{n_{1}-1}^{\ell_{1}}.
\end{align*}
Let also
$\phi^{\ell_{2}}_{n_{2}},\,n_{2}=-\ell_{2},-\ell_{2}+1,\ldots,\ell_{2}+1,\ell_{2}$
be an analogous base in the unitary space
$V_{\ell_{2}},\ell_{2}=0,\frac12,1,\frac32,\ldots$ of irreducible
representation $\mathcal{T}_{\ell_{2}}$ of another copy of the
group $\SU(2)$, generated by $G_{1},G_{2},G_{3}$, with similar
relations
\begin{align*}
W_{0}\phi_{n_{2}}^{\ell_{2}}&=n_{2}\phi_{n_{2}}^{\ell_{2}},\;
W_{+}\phi_{n_{2}}^{\ell_{2}}=-\sqrt{(\ell_{2}-n_{2})(\ell_{2}+n_{2}+1)}\psi_{n_{2}+1}^{\ell_{2}},\\
W_{-}\phi_{n_{2}}^{\ell_{2}}&=-\sqrt{(\ell_{2}+n_{2})(\ell_{2}-n_{2}+1)}\phi_{n_{2}-1}^{\ell_{2}}.
\end{align*}
Here we identify operators $T_{\pm},T_{0}$ and $W_{\pm},W_{0}$
with their restrictions onto $U_{\ell_{1}}$ and $V_{\ell_{2}}$. An
every irreducible representation of $\SO(4)$ is isomorphic to the
product $\mathcal{T}_{\ell_{1}}\otimes\mathcal{T}_{\ell_{2}}$,
where $\ell_{1},\ell_{2}$ are simultaneously integer or
half-integer numbers.

Since the group $K\cong\SO(2)$ is one-dimensional and its algebra
is generated by the element
$X^{s}_{23}=L_{1}+G_{1}=i(T_{0}+W_{0})$, subspaces
$$
\mathcal{T}'_{(\ell_{1},\ell_{2})}\subset\mathcal{T}_{(\ell_{1},\ell_{2})}:=
\mathcal{T}_{\ell_{1}}\otimes\mathcal{T}_{\ell_{2}}
$$
of decomposition (\ref{l2DirectExpansionK}) for the group $\SO(4)$
and $k=(\ell_{1},\ell_{2})$ are generated by vectors
\begin{equation}\label{TBase}
\chi_{j}^{\ell}:=\psi_{j}^{\ell_{1}}\otimes\phi_{-j}^{\ell_{2}},\,\ell=(\ell_{1},\ell_{2}),
-\min(\ell_{1},\ell_{2})\leqslant
j\leqslant\min(\ell_{1},\ell_{2}).
\end{equation}
The dimension of $\mathcal{T}'_{(\ell_{1},\ell_{2})}$ equals
$$2\min(\ell_{1},\ell_{2})+1=\ell_{1}+\ell_{2}-|\ell_{1}-\ell_{2}|+1.$$
One should find all common eigenvectors of operators
$D_{0}^{2},D_{1},D_{2}$ and optionally $D_{3}$ in the space
$\mathcal{T}'_{(\ell_{1},\ell_{2})}$.

Evidently, eigenvectors of the operator
$D_{0}^{2}=-(T_{0}-W_{0})^{2}$ are
\begin{equation}\label{ProBase1}
\chi_{0}^{\ell},\,c_{+}\chi_{j}^{\ell}+c_{-}\chi_{-j}^{\ell},\,j=1,2,\ldots,
\min(\ell_{1},\ell_{2}),
\end{equation}
if $\ell_{1},\ell_{2}$ are integer and
\begin{equation}\label{ProBase2}
c_{+}\chi_{j}^{\ell}+c_{-}\chi_{-j}^{\ell},\,j=\frac12,\frac32,\ldots,\min(\ell_{1},\ell_{2}),
\end{equation}
if $\ell_{1},\ell_{2}$ are half-integer. The corresponding
eigenvalues are $-4j^{2}$.

Since
\begin{align*}
D_{1}&=-\frac12\left(\{T_{+},T_{-}\}+\{W_{+},W_{-}\}\right)-T_{+}W_{-}-T_{-}W_{+},\\
D_{2}&=-\frac12\left(\{T_{+},T_{-}\}+\{W_{+},W_{-}\}\right)+T_{+}W_{-}+T_{-}W_{+},
\end{align*}
one should choose eigenvectors of the operators
$T_{+}W_{-}+T_{-}W_{+}$ and $\{T_{+},T_{-}\}+\{W_{+},W_{-}\}$ from
(\ref{ProBase1}) and (\ref{ProBase2}).

The base (\ref{TBase}) consists of eigenvectors of the operator
$$
\{T_{+},T_{-}\}+\{W_{+},W_{-}\}=-(T_{0}-W_{0})^{2}-D_{0}^{2}-D_{1}-D_{2}.
$$
In fact
\begin{align*}
\left(\{T_{+},T_{-}\}+\{W_{+},W_{-}\}\right)\chi_{j}^{\ell}
=2(\ell_{1}(\ell_{1}+1)+\ell_{2}(\ell_{2}+1)-2j^{2})\chi_{j}^{\ell},
\end{align*}
therefore it is enough to choose eigenvectors of the operators
$T_{+}W_{-}+T_{-}W_{+}$ from (\ref{ProBase1}) and
(\ref{ProBase2}). Since
$$
\left(T_{+}W_{-}+T_{-}W_{+}\right)\chi_{0}^{\ell}=
\sqrt{\ell_{1}\ell_{2}(\ell_{1}+1)(\ell_{2}+1)}(\chi_{1}^{\ell}+\chi_{-1}^{\ell}),
$$
one gets common eigenvectors
$\chi_{0}^{(\ell_{1},0)},\chi_{0}^{(0,\ell_{2})}$, where
$\ell_{1}$ and $\ell_{2}$ are integer.

Let $\varepsilon_{\ell}:=1$ if $\ell_{1},\ell_{2}$ are integer and
$\varepsilon_{\ell}:=\frac12$ if $\ell_{1},\ell_{2}$ are
half-integer. Since for
$j=\varepsilon_{\ell},\varepsilon_{\ell}+1,\ldots,\min(\ell_{1},\ell_{2})$
it holds
\begin{align*}
(T_{+}W_{-}&+T_{-}W_{+})(c_{+}\chi_{j}^{\ell}+c_{-}\chi_{-j}^{\ell})=
\sqrt{(\ell_{1}-j)(\ell_{1}+j+1)(\ell_{2}-j)(\ell_{2}+j+1)}(c_{+}\chi_{j+1}^{\ell}\\&+c_{-}
\chi_{-j-1}^{\ell})+\sqrt{(\ell_{1}+j)(\ell_{1}-j+1)(\ell_{2}+j)(\ell_{2}-j+1)}(c_{+}
\chi_{j-1}^{\ell}+c_{-}\chi_{-j+1}^{\ell}),
\end{align*}
the requirement
\begin{equation*}
(T_{+}W_{-}+T_{-}W_{+})(c_{+}\chi_{j}^{\ell}+c_{-}\chi_{-j}^{\ell})\sim
(c_{+}\chi_{j}^{\ell}+c_{-}\chi_{-j}^{\ell})
\end{equation*}
implies $(\ell_{1}-j)(\ell_{1}+j+1)(\ell_{2}-j)(\ell_{2}+j+1)=0$,
that gives two cases: $\varepsilon_{\ell}\leqslant
j=\ell_{1}\leqslant\ell_{2}$ and $\varepsilon_{\ell}\leqslant
j=\ell_{2}\leqslant\ell_{1}$.

In the first case one obtains
\begin{equation*}
\sqrt{2\ell_{1}(\ell_{2}+\ell_{1})(\ell_{2}-\ell_{1}+1)}(c_{+}\chi_{\ell_{1}-1}^{\ell}+c_{-}\chi_{-\ell_{1}+1}^{\ell})
=c(c_{+}\chi_{\ell_{1}}^{\ell}+c_{-}\chi_{-\ell_{1}}^{\ell}),\,c\in\mathbb{C},
\end{equation*}
that means either $\ell_{1}-1=-\ell_{1}$ or
$\ell_{1}-1=-\ell_{1}+1=0,c_{+}+c_{-}=0$. This gives the following
eigenvectors
$\chi_{\frac12}^{(\frac12,\ell_{2})}\pm\chi_{-\frac12}^{(\frac12,\ell_{2})},
\,c=\pm(\ell_{2}+\frac12)$ and
$\chi_{1}^{(1,\ell_{2})}-\chi_{-1}^{(1,\ell_{2})},\,c=0$.

In the second case one similarly gets eigenvectors of the operator
$T_{+}W_{-}+T_{-}W_{+}$ in the form:
\begin{align*}
\left(T_{+}W_{-}+T_{-}W_{+}\right)(\chi_{\frac12}^{(\ell_{1},\frac12)}\pm\chi_{-\frac12}^{(\ell_{1},\frac12)})
&=\pm(\ell_{1}+\frac12)(\chi_{\frac12}^{(\ell_{1},\frac12)}\pm\chi_{-\frac12}^{(\ell_{1},\frac12)}),\\
\left(T_{+}W_{-}+T_{-}W_{+}\right)(\chi_{1}^{(\ell_{1},1)}-\chi_{-1}^{(\ell_{1},1)})
&=0.
\end{align*}

Only two first vectors
$\chi_{0}^{(\ell_{1},0)},\chi_{0}^{(0,\ell_{2})}$ found above are
eigenvectors of the operator $D_{3}=\ii(T_{-}W_{+}-T_{+}W_{-})$.

This consideration is summarized in the following theorem.
\begin{theore}\label{eigenvectorsS3}
In the space $\mathcal{L}^2\left(\SO(4),\SO(2),d\eta_{s}\right)$
there are eight partially overlapping series of common
eigenvectors for operators $D_{0}^{2},D_{1}$ and $D_{2}:$
\begin{enumerate}
\item $D_{0}\chi_{0}^{(\ell_{1},0)}=D_{3}\chi_{0}^{(\ell_{1},0)}=0,
D_{1}\chi_{0}^{(\ell_{1},0)}=D_{2}\chi_{0}^{(\ell_{1},0)}
=-\ell_{1}(\ell_{1}+1)\chi_{0}^{(\ell_{1},0)},\ell_{1}=0,1,2,\ldots;$
\item $D_{0}\chi_{0}^{(0,\ell_{2})}=D_{3}\chi_{0}^{(0,\ell_{2})}=0,
D_{1}\chi_{0}^{(0,\ell_{2})}=D_{2}\chi_{0}^{(0,\ell_{2})}
=-\ell_{2}(\ell_{2}+1)\chi_{0}^{(0,\ell_{2})},\ell_{2}=0,1,2,\ldots;$
\item $D_{0}^{2}(\chi_{\frac12}^{(\frac12,\ell_{2})}+\chi_{-\frac12}^{(\frac12,\ell_{2})})
=-(\chi_{\frac12}^{(\frac12,\ell_{2})}+\chi_{-\frac12}^{(\frac12,\ell_{2})}),\\
D_{1}(\chi_{\frac12}^{(\frac12,\ell_{2})}+\chi_{-\frac12}^{(\frac12,\ell_{2})})
=-(\ell_{2}^{2}+2\ell_{2}+\frac34)(\chi_{\frac12}^{(\frac12,\ell_{2})}+\chi_{-\frac12}^{(\frac12,\ell_{2})}),\\
D_{2}(\chi_{\frac12}^{(\frac12,\ell_{2})}+\chi_{-\frac12}^{(\frac12,\ell_{2})})
=-(\ell_{2}^{2}-\frac14)(\chi_{\frac12}^{(\frac12,\ell_{2})}+\chi_{-\frac12}^{(\frac12,\ell_{2})}),\\
D_{3}(\chi_{\frac12}^{(\frac12,\ell_{2})}+\chi_{-\frac12}^{(\frac12,\ell_{2})})
=-\ii(\ell_{2}+\frac12)(\chi_{\frac12}^{(\frac12,\ell_{2})}-\chi_{-\frac12}^{(\frac12,\ell_{2})}),\,
\ell_{2}=\frac12,\frac32,\ldots;$
\item $D_{0}^{2}(\chi_{\frac12}^{(\ell_{1},\frac12)}+\chi_{-\frac12}^{(\ell_{1},\frac12)})
=-(\chi_{\frac12}^{(\ell_{1},\frac12)}+\chi_{-\frac12}^{(\ell_{1},\frac12)}),\\
D_{1}(\chi_{\frac12}^{(\ell_{1},\frac12)}+\chi_{-\frac12}^{(\ell_{1},\frac12)})
=-(\ell_{1}^{2}+2\ell_{1}+\frac34)(\chi_{\frac12}^{(\ell_{1},\frac12)}+\chi_{-\frac12}^{(\ell_{1},\frac12)}),\\
D_{2}(\chi_{\frac12}^{(\ell_{1},\frac12)}+\chi_{-\frac12}^{(\ell_{1},\frac12)})
=-(\ell_{1}^{2}-\frac14)(\chi_{\frac12}^{(\ell_{1},\frac12)}+\chi_{-\frac12}^{(\ell_{1},\frac12)}),\\
D_{3}(\chi_{\frac12}^{(\ell_{1},\frac12)}+\chi_{-\frac12}^{(\ell_{1},\frac12)})
=-\ii(\ell_{1}+\frac12)(\chi_{\frac12}^{(\ell_{1},\frac12)}-\chi_{-\frac12}^{(\ell_{1},\frac12)}),\,
\ell_{1}=\frac12,\frac32,\ldots;$
\item $D_{0}^{2}(\chi_{\frac12}^{(\frac12,\ell_{2})}-\chi_{-\frac12}^{(\frac12,\ell_{2})})
=-(\chi_{\frac12}^{(\frac12,\ell_{2})}-\chi_{-\frac12}^{(\frac12,\ell_{2})}),\\
D_{1}(\chi_{\frac12}^{(\frac12,\ell_{2})}-\chi_{-\frac12}^{(\frac12,\ell_{2})})
=-(\ell_{2}^{2}-\frac14)(\chi_{\frac12}^{(\frac12,\ell_{2})}-\chi_{-\frac12}^{(\frac12,\ell_{2})}),\\
D_{2}(\chi_{\frac12}^{(\frac12,\ell_{2})}-\chi_{-\frac12}^{(\frac12,\ell_{2})})
=-(\ell_{2}^{2}+2\ell_{2}+\frac34)(\chi_{\frac12}^{(\frac12,\ell_{2})}-\chi_{-\frac12}^{(\frac12,\ell_{2})}),\\
D_{3}(\chi_{\frac12}^{(\frac12,\ell_{2})}-\chi_{-\frac12}^{(\frac12,\ell_{2})})
=\ii(\ell_{2}+\frac12)(\chi_{\frac12}^{(\frac12,\ell_{2})}+\chi_{-\frac12}^{(\frac12,\ell_{2})}),\,
\ell_{2}=\frac12,\frac32,\ldots;$
\item $D_{0}^{2}(\chi_{\frac12}^{(\ell_{1},\frac12)}-\chi_{-\frac12}^{(\ell_{1},\frac12)})
=-(\chi_{\frac12}^{(\ell_{1},\frac12)}-\chi_{-\frac12}^{(\ell_{1},\frac12)}),\\
D_{1}(\chi_{\frac12}^{(\ell_{1},\frac12)}-\chi_{-\frac12}^{(\ell_{1},\frac12)})
=-(\ell_{1}^{2}-\frac14)(\chi_{\frac12}^{(\ell_{1},\frac12)}-\chi_{-\frac12}^{(\ell_{1},\frac12)}),\\
D_{2}(\chi_{\frac12}^{(\ell_{1},\frac12)}-\chi_{-\frac12}^{(\ell_{1},\frac12)})
=-(\ell_{1}^{2}+2\ell_{1}+\frac34)(\chi_{\frac12}^{(\ell_{1},\frac12)}-\chi_{-\frac12}^{(\ell_{1},\frac12)}),\\
D_{3}(\chi_{\frac12}^{(\ell_{1},\frac12)}-\chi_{-\frac12}^{(\ell_{1},\frac12)})
=\ii(\ell_{1}+\frac12)(\chi_{\frac12}^{(\ell_{1},\frac12)}+\chi_{-\frac12}^{(\ell_{1},\frac12)}),\,
\ell_{1}=\frac12,\frac32,\ldots;$
\item $D_{0}^{2}(\chi_{1}^{(1,\ell_{2})}-\chi_{-1}^{(1,\ell_{2})})
=-4(\chi_{1}^{(1,\ell_{2})}-\chi_{-1}^{(1,\ell_{2})}),\\
D_{1}(\chi_{1}^{(1,\ell_{2})}-\chi_{-1}^{(1,\ell_{2})})=D_{2}(\chi_{1}^{(1,\ell_{2})}-\chi_{-1}^{(1,\ell_{2})})
=-\ell_{2}(\ell_{2}+1)(\chi_{1}^{(1,\ell_{2})}-\chi_{-1}^{(1,\ell_{2})}),\\
D_{3}(\chi_{1}^{(1,\ell_{2})}-\chi_{-1}^{(1,\ell_{2})})=2\sqrt{2\ell_{2}(\ell_{2}+1)}\ii\chi_{0}^{(1,\ell_{2})},\,
\ell_{2}=1,2,\ldots;$
\item
$D_{0}^{2}(\chi_{1}^{(\ell_{1},1)}-\chi_{-1}^{(\ell_{1},1)})
=-4(\chi_{1}^{(\ell_{1},1)}-\chi_{-1}^{(\ell_{1},1)}),\\
D_{1}(\chi_{1}^{(\ell_{1},1)}-\chi_{-1}^{(\ell_{1},1)})=
D_{2}(\chi_{1}^{(\ell_{1},1)}-\chi_{-1}^{(\ell_{1},1)})
=-\ell_{1}(\ell_{1}+1)(\chi_{1}^{(\ell_{1},1)}-\chi_{-1}^{(\ell_{1},1)}),\\
D_{3}(\chi_{1}^{(\ell_{1},1)}-\chi_{-1}^{(\ell_{1},1)})=2\sqrt{2\ell_{1}(\ell_{1}+1)}\ii\chi_{0}^{(\ell_{1},1)},\,
\ell_{1}=1,2,\ldots$
\end{enumerate}
Only the first and the second vectors are also eigenvectors for
the operator $D_{3}$.
\end{theore}

Seeking an eigenfunction of the operator $\widehat{H}_s$ in the
form $f(r)\psi$, where $\psi$ is some vector from theorem
\ref{eigenvectorsS3}, one gets spectral equation for the two-body
problem on the sphere $\mathbf{S}^{3}$ in the form
\begin{equation}\label{GeneralFormS3}
-\frac{(1+r^{2})^{3}}{8mR^{2}r^{2}}\frac{\partial}{\partial
r}\left(\frac{r^{2}}{1+r^{2}}f'\right)+
\left(\frac1{mR^{2}}\left(\frac{a}{r^{2}}+b+cr^{2}\right)+U(r)-E\right)f=0.
\end{equation}

The first and the second case of theorem \ref{eigenvectorsS3}
correspond to arbitrary particle masses $m_{1},m_{2}$ and
equalities
$$
a=c=\frac{\ell(\ell+1)}8,
b=\frac{\ell(\ell+1)}4,\,\ell=0,1,2,\ldots
$$
In other cases both particle masses equals $2m$ and it holds
\begin{align*}
a&=\frac18(\ell^{2}-\frac14),
b=\frac14(\ell^{2}+\ell+\frac34),c=\frac18(\ell^{2}+2\ell+\frac34),\,\ell=\frac12,\frac32,
\frac52,\ldots\quad \text{in cases 3 and 4};\\
a&=\frac18(\ell^{2}+2\ell+\frac34),
b=\frac14(\ell^{2}+\ell+\frac34),c=\frac18(\ell^{2}-\frac14),\,\ell=\frac12,\frac32,\frac52,\ldots\quad
\text{in cases 5 and 6};
\\ a&=c=\frac{\ell(\ell+1)}8,
b=\frac{\ell^{2}+\ell+2}4,\,\ell=1,2,3\ldots\quad \text{in cases 7
and 8}.
\end{align*}

Note that the spectral one-particle equation for the radial
component $\psi(r)$ of an eigenfunction has the form
\begin{equation}\lb{onepart}
-\frac{(1+r^{2})^{3}}{8mR^{2}r^{2}}\frac{\partial}{\partial
r}\left(\frac{r^{2}}{1+r^{2}}f'\right)+\left(\frac{l(l+1)}{8mR^2}
(r^{-2}+2+r^2)+U-E\right)f(r)=0,\quad l=0,1,2\dots
\end{equation}
Therefore energy levels can be exactly found from equation
(\ref{GeneralFormS3}) for $a=c$ iff they can be found for the
one-particle problem with the same potential.

Usually, the spectrum of an ordinary differential operator can be
exactly found if the corresponding equation can be solved in
elementary functions or it can be reduced to the hypergeometric
equation (or its limiting cases). The hypergeometric equation is a
particular case of the Riemann equation, while the latter can be
reduced to the former by well-known linear transformations of a
dependent variable \cite{Go}.

The equation
\begin{eqnarray}\lb{gen}
-\frac{(1+r^{2})^{3}}{8mR^{2}r^{2}}\frac{\partial}{\partial
r}\left(\frac{r^{2}}{1+r^{2}}f'\right)+\left(\eta r^{-2}+\nu
r^2-E\right)f(r)=0, \quad \eta,\nu=\const
\end{eqnarray}
is the Riemannian one w.r.t.\ the independent variable $\xi=r^2$.
For $\eta,\nu>0$ the corresponding differential operator admits
the Friedrichs self-adjoint extension \cite{Shch2}, and the energy
levels are
\begin{align}\lb{ener_gen}
E_k&=\mu\left[k(k+1)-\frac58+(2k+1)\left(
\sqrt{\frac1{16}+\frac{\eta}{\mu}}+\sqrt{\frac1{16}+\frac{\nu}{\mu}}\right)
\right.\nonumber \\ &+ \left.
2\sqrt{\frac1{16}+\frac{\eta}{\mu}}\sqrt{\frac1{16}+\frac{\nu}{\mu}}\right],\;
\mu=\frac1{2mR^2}.\end{align}

By the obvious change of variables in (\ref{ener_gen}) one can
easily find energy levels for the equation (\ref{GeneralFormS3})
with the potential $U=\alpha r^{-2}+\beta
r^2,\;\alpha,\beta\geqslant 0$.

For the sphere ${\bf S}^n$ the analogs of the Coulomb and
oscillator potentials are \cite{Kil2},\cite{Lib}
\begin{equation}\lb{BetrandPotentials}
U_q=\frac{\gamma}{2R}\left(r-\frac1r\right),\quad
U_o=\frac{2\omega^2R^2r^2} {(1-r^2)^2}.
\end{equation}
All trajectories of a classical one-particle motion are closed for
these potentials. Equation (\ref{onepart}) for potentials
(\ref{BetrandPotentials}) can be reduced to the Riemann one by
changing the independent variable $r\to u=(1-r^2)/r$ for the
Coulomb potential and $r\to v=u^2$ for the oscillator one.

However the coefficients of equation (\ref{gen}) for these
potentials are rational in the independent variables $u$ and $v$
only for $\eta=\nu$. This way therefore allows us to reduce
equation (\ref{GeneralFormS3}) with potentials $U=U_q$ and $U=U_o$
to the Riemann equation only for $a=c$, i.e.\ in cases 1,2,7,8 of
theorem \ref{eigenvectorsS3}.

Theorem \ref{th3} implies the self-adjointness of the operator
$\widehat{H}_s$ with $U=U_q$ for any $n\geqslant 2$. For the
operator $\widehat{H}_s$ with $U=U_o$ we use the Friedrichs
self-adjoint extension. Energy levels for equation (\ref{onepart})
with $U=U_q$ are (see, for example, \cite{Schr}, \cite{Inf2}):
$$E_k=-\frac1{2mR^2}+\frac{(k+l)^2}{2mR^2}-\frac{m\gamma^2}{2(k+l)^2},\;
k=1,2,3\dots.
$$
Changing coefficients one can find energy levels for equation
(\ref{GeneralFormS3}) in cases 1,2,7,8 of theorem
\ref{eigenvectorsS3} with $U=U_q$
\begin{align*}
E_{k}&=\frac1{mR^{2}}\left(\frac12\left(k^{2}-k+1\right)-\frac{3}4
+2c+b+\frac{2k-1}4\sqrt{1+32c}\right)\\
&-\frac{2m\gamma^{2}}{\left(\sqrt{1+32c}+2k-1\right)^{2}},\,k\in\mathbb{N}.
\end{align*}
Similarly, the formula
$$E_k=-\frac1{2mR^2}\left(\frac34-\left(2k+l+\frac32\right)^2\right)+\frac{\omega(2k+l+
\frac32)}{\sqrt{m}}\sqrt{1+\frac1{4\omega^2R^4m}},\; k=0,1,2\dots,
$$
for energy levels of equation (\ref{onepart}) with $U=U_o$ implies
the following energy levels for equation (\ref{GeneralFormS3}) in
cases 1,2,7,8 of theorem \ref{eigenvectorsS3} with $U=U_o$
\begin{align*}\begin{split}
E_{k}&=\frac1{8mR^{2}}\left(\left(4k+2+\sqrt{1+32c}\right)^{2}-
16c+8b-3\right)
\\&+\frac{\omega}{2\sqrt{m}}\left(4k+2+\sqrt{1+32c}\right)
\sqrt{1+\frac1{4R^{4}m^{2}}},\,k=0,1,2,\ldots\end{split}
\end{align*}

\section{Reduction of cotangent bundles of homogeneous
manifolds}\label{ClassReduc}

Results of this section will be used below for the two-body
problem on constant curvature spaces.

Recall that an action of a Lie group $\Gamma$ on a smooth manifold
$M$ is {\it proper}, if for the map $\Gamma\times M\rightarrow
M\times M,\,(g,x)\rightarrow (gx,x)$ preimages of all compact sets
are compact. If additionally this action is free, then the orbit
space $\widetilde{M}:=\Gamma\backslash M$ is endowed with a
structure of a smooth manifold such that the canonical projection
$\pi_{1}\label{pi14}:\,M_{c}\rightarrow \widetilde{M}_{c}$ is a
smooth map \cite{Gui}.

Let $\Gamma$ be a Lie group with the Lie algebra $\mathfrak{g}$,
$\Gamma_0\subset\Gamma$ be some subgroup with the Lie algebra
$\mathfrak{g}_0\subset\mathfrak{g}$, acting on $\Gamma$ by right
shifts. Denote by $M=T^*\Gamma_1$ the cotangent bundle of the
homogeneous space $\Gamma_1=\Gamma/\Gamma_0$ endowed with the
standard symplectic structure. The standard $\Gamma$-action on $M$
by left shifts is Poisson \cite{Arn}. Let
$\Phi:M\to\mathfrak{g}^*$ be the corresponding momentum map and
$H$ be a $\Gamma$-invariant function on $M$. Consider the method
of the Hamiltonian reduction \cite{MW} for the Hamiltonian
dynamical system with the Hamilton function $H$ on the space $M$.
It is well known \cite{Arn} that for $\Gamma_0=\{e\}$ the reduced
phase space is symplectomorphic to the coadjoint orbit of the
group $\Gamma$, endowed with the Kirillov form. Theorem \ref{th6}
below is a generalization of this fact.

Let $\Gamma_{\beta_0}$ be the stationary subgroup of the group
$\Gamma$ w.r.t.\ some point $\beta_{0}\in\mathfrak{g}^{*}$ and the
$\Ad^{*}_{\Gamma}$-action, $\mathcal{O}_{\beta_0}$ be the orbit of
$\Ad^{*}_{\Gamma}$-action, containing the point
$\beta_0\in\mathfrak{g}^*$. Denote
$\mathcal{O}^{\prime}_{\beta_0}:=\left\{\left.\beta\in
\mathcal{O}_{\beta_0}\right|\left.\beta
\right|_{\mathfrak{g}_0}=0\right\}$. Obviously,
$\Ad^*_{\Gamma_0}\mathcal{O}^{\prime}_{\beta_0}=\mathcal{O}^{\prime}_{
\beta_0} $. Let $\widetilde{\mathcal{O}}_{\beta_0}=
\left.\mathcal{O}_{\beta_0}^{\prime}\right/\Ad^*_{ \Gamma_0}$ and
$\pi:\mathcal{O}^{\prime}_{\beta_0}\to\widetilde{\mathcal{O}}_{\beta_0}$
be the canonical projection. Let $\omega$ be the restriction of
the Kirillov form onto $\mathcal{O}^{\prime}_{\beta_0}$. Therefore
for elements $X,Y\in
T_{\beta}\mathcal{O}^{\prime}_{\beta_0},\:\beta\in\mathcal{O}^{\prime}_{\beta_0}$
of the form
$$X=\left.\frac{d}{dt}\right|_{t=0}\Ad^*_{\exp(tX^{\prime})}\beta,\;
  Y=\left.\frac{d}{dt}\right|_{t=0}\Ad^*_{\exp(tY^{\prime})}\beta,\;
  X^{\prime},Y^{\prime}\in\mathfrak{g},$$
one has $\omega(X,Y)=\beta\left([X^{\prime},Y^{\prime}]\right)$.
Due to
$\left.\Ad^*_{\exp(tX^{\prime})}\beta\right|_{\mathfrak{g}_0}=0$,
it holds
$$\beta\left([X^{\prime},Y_0^{\prime}]\right)=
\left.\frac{d}{dt}\right|_{t=0}\Ad^*_{\exp(tX^{\prime})}\beta(
Y^{\prime}_0)=0$$ for any element $Y_0^{\prime}\in\mathfrak{g}_0$.
It means that the formula
$\widetilde{\omega}(\widetilde{X},\widetilde{Y})=\omega(d\pi^{-1}\widetilde{X},d\pi^{-1}
\widetilde{Y})$ defines the 2-form $\widetilde{\omega}$ on
$T\widetilde{\mathcal{O}}_{\beta_0}$ for $\widetilde{X}\in
T_{\pi\beta}\widetilde{\mathcal{O}}_{\beta_0},\: \widetilde{Y}\in
T_{\pi\beta}\widetilde{\mathcal{O}}_{\beta_0}$.

\begin{theore}\label{th6}
Suppose that the orbit $\mathcal{O}_{\beta_0}$ is transversal to
the subspace $\Ann\mathfrak{g}_{0}\subset\mathfrak{g}^{*}$ and
therefore the set $\mathcal{O}^{\prime}_{\beta_0}$ is a
submanifold of the orbit $\mathcal{O}_{\beta_0}$. Let also the
$\Ad_{\Gamma_{0}}^{*}$-action on the space
$\mathcal{O}^{\prime}_{\beta_0}$ be free and proper. Then the
reduced phase space $\widetilde{M}_{\beta_0}$, corresponding to
the value $\beta_0$ of the momentum map, is symplectomorphic to
the symplectic space
$\left(\widetilde{\mathcal{O}}_{\beta_0},\:\widetilde{\omega}\right)$.
\end{theore}
\begin{proof}
Consider a point $x\in M_{\beta_0}$ of the level set
$$M_{\beta_0}:=\Phi^{-1}(\beta_0)\subset M$$ for the momentum map as the orbit
$\mathcal{O}_{x^{\prime}}$ of some point $x^{\prime}=
(\gamma,p)\in T^*\Gamma,\:\gamma\in\Gamma,\: p\in
T^*_{\gamma}\Gamma$ under right $\Gamma_0$-shifts on $T^*\Gamma$.
To avoid cumbersome notations we preserve symbols $L_{\gamma_1}$
and $R_{\gamma_1}$ respectively for the left $(\gamma,p)\to
(\gamma_1\gamma,L^*_{\gamma_1^{-1}}p)$ and the right
$(\gamma,p)\to (\gamma\gamma_1,R^*_{\gamma_1^{-1}}p)$ actions of
an element $\gamma_1\in\Gamma$ on $T^*\Gamma$. Due to the
definition of the momentum map \cite{Arn} for a vector
$$
X=\left.\frac{d}{dt}\right|_{t=0}L_{\exp(tX^{\prime})}\gamma,\;
X^{\prime}\in\mathfrak{g},\: X\in T_{\gamma}\Gamma
$$
it holds $p(X)=\beta_0(X^{\prime})$, i.e.\
$p=R^*_{\gamma^{-1}}\beta_0$. If additionally $X^{\prime}\in
\Ad_{\gamma}\mathfrak{g}_0$, then $X\in
d\pi_1\left(T_{x^{\prime}}\mathcal{O}_{x^{\prime}}\right)$, where
$\pi_1:\: T^*\Gamma\to\Gamma$ is the canonical projection, and
$p(X)=0$. Thus one gets
$\left.\Ad^*_{\gamma}\beta_0\right|_{\mathfrak{g}_0}=0$.

Denote $\mathcal{O}=\left\{x^{\prime}=(\gamma,p)\in
\left.T^*\Gamma\right|\;\left.\Ad^*_{\gamma}\beta_0\right|_{\mathfrak{g}_0}=0,
p=R^*_{\gamma^{-1}}\beta_0\right\}$. Due to the theorem
assumptions the set $\mathcal{O}$ is a submanifold of
$T^{*}\Gamma$. An action of a Lie subgroup on the whole Lie group
(or its submanifolds) by shifts is always proper. Therefore the
quotient manifold $\mathcal{O}/\Gamma_{0}$ coincides with the set
$M_{\beta_0}$, which is therefore a submanifold of the space $M$.
Let $\tau:\quad\mathcal{O}\to\mathfrak{g}^*=T_e^*\Gamma$ be a map
defined by the formula
$\tau(\gamma,p)=L^*_{\gamma}p=\Ad^{*}_{\gamma}\beta_{0}$. The
following diagram is commutative \cite{Arn}
$$\begin{CD}
T^*\Gamma @>L_{\gamma^{-1}}>> T^*\Gamma \\ @V\Phi VV @VV\Phi V
\\ \mathfrak{g}^* @>\Ad^*_{\gamma}>> \mathfrak{g}^*
\end{CD}$$
Consequently two points of the manifold $\mathcal{O}$ are mapped
by $\tau$ into one point iff they lie on one
$\Gamma_{\beta_0}$-orbit w.r.t.\ left $\Gamma_{\beta_0}$-shifts on
the manifold $\mathcal{O}$. By definition of $\mathcal{O}$ it
holds $\tau(\mathcal{O})=\mathcal{O}^{\prime}_{\beta_0}$ and
therefore $\tau$ is the quotient map
$\mathcal{O}\to\Gamma_{\beta_0}\backslash\mathcal{O}=\mathcal{O}^{\prime}_{\beta_0}$.

The point $(\gamma,p)$ is mapped by $\tau$ into
$\Ad_{\gamma}^*\beta_0$, so the point $R_{\gamma_0}(\gamma,p)$ is
mapped into $\Ad^*_{\gamma\gamma_0}\beta_0=\Ad^*_{\gamma_0}\circ
\Ad^*_{\gamma}\beta_0$. Thus $\Gamma_0$-orbits in
$\mathcal{\mathcal{O}}$ w.r.t.\ right shifts are mapped into
$\Ad^{*}_{\Gamma_0}$-orbits in $\mathcal{O}^{\prime}_{\beta_0}$.

By definition the $\Ad^{*}_{\Gamma_0}$-action in
$\mathcal{O}^{\prime}_{\beta_0}$ is free and proper, therefore the
intersection of $L_{\Gamma_{\beta_{0}}}$- and
$R_{\Gamma_{0}}$-orbits in $\mathcal{O}$ consists of no more than
of one point. This implies that $L_{\Gamma_{\beta_{0}}}$-action on
the manifold $M_{\beta_0}=\mathcal{O}/\Gamma_{0}$ is free and the
reduced space
$\widetilde{M}_{\beta_0}:=\Gamma_{\beta_{0}}\backslash
M_{\beta_0}$ is a quotient manifold.

Hence the map $\tau$ induces the diffeomorphism
$$\phi:\:\widetilde{M}_{\beta_0}=\Gamma_{\beta_0}\backslash M_{\beta_0}=
\Gamma_{\beta_0}\backslash\left(\mathcal{O}/\Gamma_0\right)=
\left(\Gamma_{\beta_0}\backslash\mathcal{O}\right)/\Gamma_0\to
\mathcal{O}^{\prime}_{\beta_0}/
\Ad^*_{\Gamma_0}=\widetilde{\mathcal{\mathcal{O}}}_{\beta_0}.$$

Finally we need to prove that the symplectic form
$\widehat{\omega}$ on $\widetilde{M}_{\beta_0}$ is mapped by
$\phi$ into the form $-\widetilde{\omega}$. However this fact is
an easy consequence of its particular case for $\Gamma_0=\{e\}$
\cite{Arn}, the possibility to represent vectors tangent to the
space $\widetilde{M}_{\beta_0}$ via vectors tangent to
$\mathcal{O}$, and the commutativity of the following diagram
$$\begin{CD}
\mathcal{O} @>R_{\gamma_0}>> \mathcal{O} \\ @V\tau VV @VV\tau V
\\ \mathcal{O}'_{\beta_0} @>R_{\gamma_0}>> \mathcal{O}'_{\beta_0}
\end{CD}$$
for any $\gamma_0\in\Gamma_0$.
\end{proof}
The form $\widehat{\omega}$ is symplectic, therefore one gets
\begin{corollary}\label{cor1}
The form $\widetilde{\omega}$ on
$\widetilde{\mathcal{O}}_{\beta_0}$ is symplectic, i.e.\ it is
nondegenerate and closed.
\end{corollary}

\section{Hamiltonian reduction of the two-body problem on constant curvature spaces }

We adjust Poisson brackets with a symplectic structure in the
following way. Let $X_{h}$ be a Hamiltonian vector field on a
symplectic space $M$, corresponding to a Hamilton function $h$,
then
\begin{equation}\label{HamFieldDef}
dh=\omega(\cdot,X_{h})\equiv-i_{X_{h}}\omega,
\end{equation}
where $i_{X}\omega$ is the contraction of the vector field $X$ and
the symplectic form $\omega$. The Poisson brackets of functions
$f$ and $h$ on $M$ are
\begin{equation}\label{PB}
[f,h]_{P}:=-\omega(X_{f},X_{h})=-dh(X_{f})=df(X_{h}).
\end{equation}

It was noted in \cite{Shch1} that the classical two-body problem
on spaces $\bHn$ and $\bSn,n\geqslant 3$ reaches its full
generality at $n=3$, since for $n>3$ any two elements from the
space $T^*\bHn$ ($T^*\bSn$) are in some subspace $T^*{\bf
H}^3\subset T^*\bHn $ ($T^*{\bf S}^3\subset T^*\bSn$). Therefore
two particles with a central interaction will always stay in some
subspace $\bHth$ (in $\bSth$). Below we consider the case $n=3$.

\subsection{Two body problem on ${\bf S^3}$}\lb{S3Reduction}

Let the space $M=T^*Q_s$ is endowed with the standard symplectic
structure of a cotangent bundle. Then due to section
\ref{S3Hamiltonian} one can represent the manifold $M$ in the form
\begin{equation}\lb{prod}
T^*\mathbb{R}_+\times T^*\left(\SO(4)/\SO(2)\right).
\end{equation}
up to a zero measure set. The symmetry group $\SO(4)$ acts only
onto the second factor of the product (\ref{prod}), therefore the
construction from section \ref{ClassReduc} can be easily
generalize for the case under consideration. The reduced phase
space for (\ref{prod}) is
$$\widetilde{M}_{\beta_0}=T^*\mathbb{R}_+\times \widetilde{\mathcal{O}}_{\beta_0},$$
where the space $\widetilde{\mathcal{O}}_{{\beta}_0}$ is
constructed for the groups $\Gamma=\SO(4),\;\Gamma_0=\SO(2)$ as in
section \ref{ClassReduc}.

Below we shall introduce coordinates in the space
$\widetilde{M}_{{\beta}_0}$ and express the reduced two-body
Hamilton function through these coordinates using formula
(\ref{H_0^s}).

For $n=3$ the Killing vector fields (\ref{basis_s}) are
$X^s_{12},X^s_{31},X^s_{23},Y^s_1,Y^s_2,Y^s_3$. In the present
section for simplicity we use the same notations for the
corresponding basis in $\mathfrak{so}(4)$ (omitting the
superscript "s") in accordance with (\ref{corresp}). Let
\begin{align*}
L^1&=X^{23}+Y^1, L^2=X^{31}+Y^2,L^3=X^{12}+Y^3,
\\ G^1&=X^{23}-Y^1,G^2=X^{31}-Y^2,G^3=X^{12}-Y^3
\end{align*}
be the base in $\mathfrak{so}^{*}(4)$, dual to (\ref{NewBase}).
Let also
\begin{equation}\lb{arb_point} {\bf
p}=p_1X^{23}+p_2X^{31}+p_3X^{12}+p_4Y^1+p_5Y^2+p_6Y^3=
\sum\limits_{i=1}^3\left(u_iL^i+v_iG^i\right)
\end{equation}
be an arbitrary element from the space $\mathfrak{so}^*(4)$.

In order to avoid cumbersome calculations, similar to calculations
in section \ref{S3Hamiltonian}, we pass from the quantum case to
the classical one changing a filtered operator algebra by the
corresponding graded one. In particular, commutator relations turn
into Poisson brackets.

Formulas (\ref{delta}) and (\ref{H_0^s}) lead to the following
expression
$$H_s=\frac{(1+r^2)^2}{8mR^2}p_r^2+\frac1a p_4^2 + \frac12A_s\left(p_2^2+p_3^2\right)
+ \frac12C_s\left(p_5^2+p_6^2\right)-
B_s\left(p_3p_5-p_2p_6\right) + U(r)
$$
for the classical Hamilton function, where $p_r$ is the momentum,
corresponding to the coordinate $r$.

Expressions
$$
\mathcal{P}_{0}:=p_{4},\,\mathcal{P}_{1}:=p_5^2+p_6^2,\,\mathcal{P}_{2}:=p_2^2+p_3^2,
\,\mathcal{P}_{3}:=-p_3p_5+p_2p_6
$$
correspond to $\SO(4)$-invariant functions on the space
$T^*\left(\SO(4)/\SO(2)\right)$. The substitution
$D_{k}\to\mathcal{P}_{k}$ and the subsequent rejection of summands
with a degree less than $\deg D_{k}+\deg D_{j}-1$ transform
commutator relations $[D_{k},D_{j}]$ (see (\ref{CommRelations}))
into Poisson brackets
$\left[\mathcal{P}_{k},\mathcal{P}_{j}\right]_{P}$. Thus one gets
\begin{align}\begin{split}
[\mathcal{P}_{0},\mathcal{P}_{1}]_{P}&=-2\mathcal{P}_{3},\,[\mathcal{P}_{0},
\mathcal{P}_{2}]_{P}=2\mathcal{P}_{3},\,
[\mathcal{P}_{0},\mathcal{P}_{3}]_{P}=\mathcal{P}_{1}-\mathcal{P}_{2},\\
[\mathcal{P}_{1},\mathcal{P}_{2}]_{P}&=-4\mathcal{P}_{0}\mathcal{P}_{3},\,
[\mathcal{P}_{1},\mathcal{P}_{3}]_{P}=-2\mathcal{P}_{0}\mathcal{P}_{1},\,
[\mathcal{P}_{2},\mathcal{P}_{3}]_{P}=2\mathcal{P}_{0}\mathcal{P}_{2}.
\end{split}\lb{PoissonComS}\end{align}

Changing variables as $p_i=u_i+v_i,\; p_{3+i}=u_i-v_i,\; i=1,2,3$
we obtain the following form of the two-body Hamilton function
\begin{align*}
H_s &=\frac{(1+r^2)^2}{8mR^2}p_r^2+\frac1a\left(u_1-v_1\right)^2 +
\frac12A_s\left(\left(u_2+v_2\right)^2+\left(u_3+v_3\right)^2\right)
\\&+\frac12C_s\left(\left(u_2-v_2\right)^2+\left(u_3-v_3\right)^2\right)
- 2B_s\left(u_2v_3-v_2u_3\right) + U(r).
\end{align*}

We now construct canonical conjugate coordinates on the space
$\widetilde{\mathcal{O}}_{\beta_0}$. Due to the special choice of
the point ${\bf x}_0$ in the submanifold $F_r$ (see section
\ref{S3Hamiltonian}) its stabilizer $K\cong\SO(2)$ is generated by
$X_{23}$. It is well known that coadjoint orbits of the group
$\SO(3)$ are two dimensional spheres. The Kirillov form on these
spheres coincides with their area forms. Therefore the orbit
$\mathcal{O}_{\beta_0}$ is a set of points (\ref{arb_point}) such
that their coordinates $u_i,\;v_i,\;i=1,2,3$ satisfy the following
equations
\begin{equation}\lb{cond}
u_1^2+u_2^2+u_3^2=\mu^2,\; v_1^2+v_2^2+v_3^2=\nu^2,
\end{equation}
where $\mu,\nu$ are nonnegative real numbers.

The subset $\mathcal{O}'_{\beta_0}\subset\mathcal{O}_{\beta_0}$
consists of elements from $\mathcal{O}_{\beta_0}$ that are
annulled by the vector $X_{23}$ and for description of the subset
$\mathcal{O}'_{\beta_0}$ one must add the condition
$p_1=u_1+v_1=0$ to equations (\ref{cond}).

Let us verify the first assumption of theorem \ref{th6}, i.e.\
whether the orbit $\mathcal{O}_{\beta_0}$ is transversal to the
subspace $\Ann X_{23}\subset\mathfrak{so}^{*}(4)$. Consider a
point $\z\in\mathcal{O}'_{\beta_0}$ with coordinates
$$(u_{1},u_{2},u_{3},v_{1},v_{2},v_{3}=-u_{1}).$$
First let $\mu,\nu>0$. A vector
$$
Z=\sum_{i=1}^{3}\left(y_{i}L^{i}+z_{i}G^{i}\right)
$$
is tangent to the orbit $\mathcal{O}_{\beta_0}$ iff
\begin{equation}\label{TransVerse}
u_{1}y_{1}+u_{2}y_{2}+u_{3}y_{3}=0,\;-u_{1}z_{1}+v_{2}z_{2}+v_{3}z_{3}=0.
\end{equation}
Since $\dim\Ann X_{23}=5$, the orbit $\mathcal{O}_{\beta_0}$ is
not transversal to the subspace $\Ann X_{23}$ at the point $\z$
iff $T_{\z}\mathcal{O}_{\beta_0}\subset\Ann X_{23}$. On the
coordinate level the latter condition means that equations
(\ref{TransVerse}) imply the equality $y_{1}+z_{1}=0$. Clearly, it
is valid only for $u_{2}=u_{3}=v_{2}=v_{3}=0,\,u_{1}\ne 0$ and
thus it holds $\mu=\nu>0$.

If $\mu>0,\nu=0$, then $u_{1}=v_{1}=v_{2}=v_{3}=0$ and a vector
$Z=y_{1}L^{1}+y_{2}L^{2}+y_{3}L^{3}$ is tangent to the orbit
$\mathcal{O}_{\beta_0}$ iff
\begin{equation}\lb{TransVerse1}
u_{2}y_{2}+u_{3}y_{3}=0.
\end{equation}
Since equation (\ref{TransVerse1}) does not restrict values of
$y_{1}$, the orbit $\mathcal{O}_{\beta_0}$ is again transversal to
the subspace $\Ann X_{23}$. The case $\mu=0,\nu>0$ is completely
similar.

Thus the orbit $\mathcal{O}_{\beta_0}$ is transversal to the
subspace $\Ann X_{23}\subset\mathfrak{so}^{*}(4)$ iff $\mu\ne\nu$.

Consider the cases $\mu\ne\nu$ and $\mu=\nu$ separately.
\begin{enumerate}
\item Let $\mu\ne\nu$.
\begin{enumerate}
\item First consider the subcase $\mu,\nu>0$. Let $u,\psi,\chi$ be
coordinates on the space $\mathcal{O}'_{\beta_0}$, defined by the
following equations
\begin{align*}
u_1&=-v_1=u,\; u_2=\sqrt{\mu^2-u^2}\sin\psi,\;
u_3=\sqrt{\mu^2-u^2}\cos\psi, \\ v_2&=\sqrt{\nu^2-u^2}\sin\chi,\;
v_3=\sqrt{\nu^2-u^2}\cos\chi,\; -\min\{\mu,\nu\}\leqslant
u\leqslant\min\{\mu,\nu\}.
\end{align*}
The restriction of the Kirillov form from $\mathcal{O}_{\beta_0}$
onto $\mathcal{O}'_{\beta_0}$ is
\begin{align}\begin{split}
\omega&=\frac1{\mu^2}\left(u_1du_2\land du_3+u_2du_3\land du_1
+u_3du_1\land du_2\right) \\ &+\frac1{\nu^2}\left(v_1dv_2\land
dv_3+v_2dv_3\land v_1 +v_3dv_1\land dv_2\right)=du\land d(\psi
-\chi).
\end{split}\lb{sym_form} \end{align}
Formulas $u\to u,\;\psi\to\psi
+\xi,\;\chi\to\chi+\xi,\:0\leqslant\xi<2\pi$ describe the
$\Ad_{K}^{*}$-action in $\mathcal{O}'_{\beta_0}$. This action is
free and proper. Therefore
$\widetilde{\mathcal{O}}_{\beta_0}=\mathcal{O}'_{\beta_0}/\Ad^{*}_{K}$
is a quotient manifold with canonical conjugate coordinates
$\phi=\psi -\chi,\;p_{\phi}=u$.

For $\mu>\nu>0$ an arbitrary $\Ad^{*}_{K}$-orbit in
$\widetilde{\mathcal{O}}_{\beta_0}$ contains a unique point with
coordinates
$$
u_{1}=u,\,u_{2}=0,\,u_{3}=\sqrt{\mu^{2}-u^{2}},\,v_{1},\,v_{2},\,v_{3}=-u
$$
such that
$$
v_{1}^{2}+v_{2}^{2}+u^{2}=\nu^{2}.
$$
This implies that the space $\widetilde{\mathcal{O}}_{\beta_0}$ is
diffeomorphic to the sphere ${\bf S}^{2}$. Similarly, for
$\nu>\mu>0$ the space $\widetilde{\mathcal{O}}_{\beta_0}$ is also
diffeomorphic to the sphere ${\bf S}^{2}$.

The coordinate system $p_{\phi},\phi$ has a singularity at the
points $p_{\phi}=\pm\min\{\mu,\nu\}$. It differs from the
coordinate system on the reduced space in \cite{Shch1}. The
reduced Hamilton function is
\begin{align*}
&\widetilde{H}_s =\frac{(1+r^2)^2}{8mR^2}p_r^2+\frac{4p_{\phi}^2}a
+\frac12A_s\left(\mu^2+\nu^2-2p_{\phi}^2+2\sqrt{\mu^2-p_{\phi}^2}
\sqrt{\nu^2-p_{\phi}^2}\cos\phi\right) \\ &+
\frac12C_s\left(\mu^2+\nu^2-2p_{\phi}^2-2\sqrt{\mu^2-p_{\phi}^2}
\sqrt{\nu^2-p_{\phi}^2}\cos\phi\right)
\\&-2B_s\sqrt{\mu^2-p_{\phi}^2}\sqrt{\nu^2-p_{\phi}^2}\sin\phi +
U(r).
\end{align*}

\item In the subcase $\mu=0,\nu>0 $ (or $\nu=0,\mu>0 $) the orbit $\mathcal{O}'_{\beta_0}$
is defined by equations $u_1=u_2=u_3=v_1=0 $. Therefore it holds
$\mathcal{O}'_{\beta_0}={\bf S}^1$ and
$\widetilde{\mathcal{O}}_{\beta_0}=\pt$. The reduced phase space
is $T^*\mathbb{R}_+$ with the reduced Hamilton function
\begin{equation}\lb{IntegrableCase}
\widetilde{H}_s=\frac{(1+r^2)^2}{8mR^2}\left(p_r^2+\frac{\nu^2}{r^2}\right)+U(r),
\end{equation}
corresponding to an integrable system.
\end{enumerate}

\item Let $\nu=\mu$. This case corresponds to particle motion
along a two dimensional sphere ${\bf S}^2\subset{\bf S}^3$ (see
\cite{Shch1}, proposition 1). Therefore one can assume
$\Gamma=\SO(3)$ and $\Gamma_{0}=\{e\}$. Obviously, the
requirements of theorem \ref{th6} are satisfied.

\begin{enumerate}
\item First consider the subcase $\beta_{0}\ne0$.
In accordance with theorem \ref{th6} the reduced phase space
$\widetilde{M}_{\beta_{0}}$ of the two-body problem is
diffeomorphic to the space
$$
T^{*}\mathbb{R}_{+}\times\mathcal{O}_{\beta_{0}},
$$
where $\mathcal{O}_{\beta_{0}}\cong{\bf
S}^2\subset\mathfrak{so}^{*}(3)$. The reduced Hamilton function
has the form
$$
\widetilde{H}_s=\frac{(1+r^2)^2}{8mR^2}p_r^2+\frac1a p_4^2 + \frac12A_sp_3^2
+ \frac12C_sp_5^2- B_sp_3p_5 + U(r).
$$
The orbit $\mathcal{O}_{\beta_{0}}$ is defined by the equation
$$
p_{3}^{2}+p_{4}^{2}+p_{5}^{2}=\beta_{0}^{2}
$$
and it holds
$$
[p_{3},p_{4}]_{P}=p_{5},\,[p_{4},p_{5}]_{P}=p_{3},\,[p_{5},p_{3}]_{P}=p_{4}.
$$

\item The last subcase $\nu=\mu=0$ corresponds to particle motion
along a common geodesic ${\bf S}^1$ (see \cite{Shch1}, proposition
2). Here $\mathcal{O}_{\beta_{0}}=\pt$ and one gets
$$
\widetilde{M}_{0}=T^*\mathbb{R}_+,\; \widetilde{H}_s =
\frac{(1+r^2)^2}{8mR^2}p_r^2+U(r).$$
\end{enumerate}
\end{enumerate}

\subsection{Two body problem on ${\bf H^3}$}

After excluding the diagonal from the space $Q_{h}=\bf
H^3\times\bf H^3$ one gets the phase space of the two-body problem
in the form
\begin{equation}
\lb{prodH} T^*I\times T^*\left(\Or_{0}(1,3)/\SO(2)\right).
\end{equation}
The symmetry group $\Or_{0}(1,3)$ acts only onto the second factor
of the product (\ref{prodH}), therefore the Hamiltonian reduction
leads to the reduced space
$$\widetilde{M}_{\beta_0}=T^*I\times
\widetilde{\mathcal{O}}_{\beta_0},
$$
where $\widetilde{\mathcal{O}}_{{\beta}_0}$ is constructed for the
groups $\Gamma=\Or_{0}(1,3),\;\Gamma_0=\SO(2)$ as in section
\ref{ClassReduc}.

Since the Lie algebra $\mathfrak{so}(1,3)$ is simple, one can not
represent $\Ad_{\Or_{0}(1,3)}^{*}$-orbits as direct products
contrary to section \ref{S3Reduction}. Nevertheless dynamic
systems on the sphere ${\bf S}^3$ and the hyperbolic space ${\bf
H}^3$ are connected by the formal substitution (see section
\ref{H3Hamiltonian} and \cite{Shch1}). This motivates the
following construction.

Let $L_1=X_{23},L_2=X_{31},L_3=X_{12},Y_1,Y_2,Y_3$ be the basis in
the Lie algebra $\mathfrak{so}(1,3)$, corresponding to Killing
vector fields (\ref{basis_h}), and $L^1,L^2,L^3,Y^1,Y^2,Y^3$ be
the dual basis in $\mathfrak{so}^*(1,3)$. Let ${\bf
p}=p_1L^1+p_2L^2+p_3L^3+p_4Y^1+p_5Y^2+p_6Y^3$ be an arbitrary
element from $\mathfrak{so}^*(1,3)$. Direct calculation shows that
the expressions
$$
I_1=p_1^2+p_2^2+p_3^2-p_4^2-p_5^2-p_6^2,\quad
I_2=p_1p_4+p_2p_5+p_3p_6
$$
are invariants of $\Ad_{\Or_{0}(1,3)}^{*}$-action.

Similarly to section \ref{S3Reduction} one gets the following
expression of the two-body Hamilton function
\begin{equation}\lb{ham_class_h}
H_h=\frac{(1-r^2)^2}{8mR^2}p_r^2+\frac1a\p_{0}^{2} +
\frac12A_h\p_{2}+\frac12C_h\p_{1}- B_h\p_{3} + U(r),\;0<r<1,
\end{equation}
where expressions
$$
\p_{0}:=p_{4},\,\p_{1}:=p_5^2+p_6^2,\,\p_{2}:=p_2^2+p_3^2,
\,\p_{3}:=-p_3p_5+p_2p_6
$$
correspond to $\Or_{0}(1,3)$-invariant functions on the space
$T^*\left(\Or_{0}(1,3)/\SO(2)\right)$. One can derive Poisson
brackets $\left[\p_{k},\p_{j}\right]_{P}$ from commutator
relations (\ref{ComRelationH}) in full analogy with the derivation
of brackets (\ref{PoissonComS}) from (\ref{CommRelations})
\begin{align*}
[\p_{0},\p_{1}]_{P}&=2\p_{3},\,[\p_{0},\p_{2}]_{P}=2\p_{3},\,
[\p_{0},\p_{3}]_{P}=\p_{1}+\p_{2},\\
[\p_{1},\p_{2}]_{P}&=-4\p_{0}\p_{3},\,
[\p_{1},\p_{3}]_{P}=-2\p_{0}\p_{1},\,
[\p_{2},\p_{3}]_{P}=2\p_{0}\p_{2}.
\end{align*}

Let $\mathcal{O}_{\beta_0}$ be an $\Ad_{\Or_{0}(1,3)}^{*}$-orbit
defined by equations $I_1=\mu,I_2=\nu\ne0,\;\mu,\nu\in\mathbb{R}$.
Therefore the subset $\mathcal{O}'_{\beta_0}$ is defined by
equations $I_1=\mu,I_2=\nu,p_1=0$. The stationary subgroup
$K\simeq\SO(2)$ of the point ${\bf x}_0\in F_r$ is generated by
the element $L_1$ and the $\Ad_{K}^{*}$-action coincides with the
simultaneous rotation in coordinate planes $(p_2,p_3)$ and
$(p_5,p_6)$. Likewise in section \ref{S3Reduction} one can verify
that for $\nu\ne0$ the orbit $\mathcal{O}_{\beta_0}$ is
transversal to $\Ann L_{1}\subset\mathfrak{so}^{*}(1,3)$ and the
first assumption of theorem \ref{th6} is valid.

\begin{enumerate}
\item Let $\nu\ne0$. The following formulas define coordinates
$p_4,\psi,\chi $ on the manifold $\mathcal{O}'_{\beta_0}$
\begin{gather}\begin{split}
p_2=u\cosh\psi\cos\chi + v\sinh\psi\sin\chi,
p_3=v\sinh\psi\cos\chi - u\cosh\psi\sin\chi, \\
p_5=v\cosh\psi\cos\chi - u\sinh\psi\sin\chi,
p_6=-u\sinh\psi\cos\chi - v\cosh\psi\sin\chi,
\end{split}\lb{impulses}\end{gather}
where $p_4,\psi\in\mathbb{R},\chi\in\mathbb{R}\;\mod\;2\pi$ and
values $u,v$ are defined by equations
\begin{equation}\lb{eq_uv} u^2-v^2=\mu+p_4^2,\quad uv=\nu.
\end{equation}
Two solutions of (\ref{eq_uv}) differ in sign and it suffice to
choose either of them. The $\Ad^{*}_{K}$-action is free, proper
and corresponds to the rotation $\chi\to\chi+\xi$. Thus theorem
\ref{th6} is applicable.

An every $\Ad^{*}_{K}$-orbit in $\mathcal{O}'_{\beta_0}$ contains
a unique point with coordinates
$$
p_{1}=p_{2}=0,\,p_{3}>0,p_{4},p_{5},p_{6}=\frac{\nu}{p_{3}}
$$
such that
$$
p_{4}^{2}+p_{5}^{2}=p_{3}^{2}-\frac{\nu^{2}}{p_{3}^{2}}-\mu.
$$
This equation defines a unique positive $p_{3}$, therefore the
space $\widetilde{\mathcal{O}}_{\beta_0}$ is diffeomorphic to the
plane $\mathbb{R}^2$ with global coordinates $p_{4},p_{5}$.

Thus for $\nu\ne0$ the reduced phase space
$\widetilde{M}_{\beta_{0}}$ of the two-body problem in ${\bf H^3}$
is diffeomorphic to the space
$$
T^{*}I\times\mathbb{R}^{2}.
$$

It is well known that $\Ad^{*}_{G}$-orbits of an arbitrary Lie
group $G$ coincides with symplectic leaves of the canonical
Poisson structure in the space $\mathfrak{g}^{*}$. Let
$\{e_i\}_{i=1}^n $ be a basis in the Lie algebra
$\mathfrak{g},\quad [e_i,e_j]=c_{ij}^k e_k $ and $\{x_i\}_{i=1}^n$
be coordinates on $g^*$, corresponding to the dual basis
$\{e^i\}_{i=1}^n $. Let also $f_1,f_2$ be arbitrary smooth
functions on $\mathfrak{g}^*$. Then their Poisson brackets has the
form
$$\left[f_1,f_2\right]_{P}=\sum\limits_{i,j,k=1}^n c^k_{ij}x_k
\frac{\partial f_1}{\partial x_i}\frac{\partial f_2}{\partial
x_j}.
$$
The choice of a sing in this equation is defined by the
correspondence between commutators of differential operators and
Poisson brackets of corresponding functions.

We shall use Poisson brackets on $\mathfrak{so}^*(1,3)$ for
construction of canonical conjugate coordinates on the space
$\widetilde{\mathcal{O}}_{\beta_0}$. Formulas
\begin{gather*}
\psi = \frac14\ln\left(\frac{(p_2-p_6)^2+(p_5+p_3)^2}
{(p_2+p_6)^2+(p_5-p_3)^2}\right), \\ \chi =
\frac12\left(\arctan\left(\frac{p_5-p_3}{p_2+p_6}\right) -
\arctan\left(\frac{p_5+p_3}{p_2-p_6}\right)\right),
\end{gather*}
$$ [L_i,L_j ] = \sum\limits_{k=1}^3\varepsilon_{ijk} L_k,\quad
 [ Y_i,Y_j ] = -\sum\limits_{k=1}^3\varepsilon_{ijk} L_k,\quad
 [ L_i,Y_j ] =\sum\limits_{k=1}^3\varepsilon_{ijk}Y_k,$$
yield the following relations
$$[p_4,\psi]_{P}=-1,\quad[p_4,\chi]_{P}=0,\quad[\psi,\chi]_{P}=0.$$
Therefore equations (\ref{HamFieldDef}) and (\ref{PB}) imply that
the symplectic structure on the space
$\widetilde{\mathcal{O}}_{\beta_0}$ is defined by $dp_4\land
d\psi$. From (\ref{impulses}) one gets
\begin{gather*}
p_2^2+p_3^2=
\frac12\left(\mu+p_4^2+\sqrt{\left(\mu+p_4^2\right)^2+4\nu^2}
\cosh 2\psi\right), \\ p_5^2+p_6^2 =
\frac12\left(-\mu-p_4^2+\sqrt{\left(\mu+p_4^2\right)^2+4\nu^2}
\cosh 2\psi\right), \\ p_3p_5-p_2p_6 =
\frac12\sqrt{\left(\mu+p_4^2\right)^2+4\nu^2}\sinh 2\psi.
\end{gather*}
Introducing the new canonical conjugate coordinates
$p_{\phi}=p_4/2,\phi=2\psi$, one gets from (\ref{ham_class_h}) the
final form of the reduced Hamilton function
\begin{align*}
&\widetilde{H}_h =\frac{(1-r^2)^2}{8mR^2}p_r^2+\frac{4p_{\phi}^2}a
+\frac12A_h\left(\frac{\mu}2+2p_{\phi}^2+2\sqrt{\left(\frac{\mu}4+p_{\phi}^2\right)^2
+\frac{\nu^2}4}\cosh\phi\right) \\ &-\frac12
C_h\left(\frac{\mu}2+2p_{\phi}^2-2\sqrt{\left(\frac{\mu}4+p_{\phi}^2\right)^2
+\frac{\nu^2}4}\cosh\phi\right)\\&-2B_h
\sqrt{\left(\frac{\mu}4+p_{\phi}^2\right)^2+\frac{\nu^2}4}\sinh\phi
+U(r).
\end{align*}

\item The case $\nu=0$ corresponds to the particle motion along
a hyperbolic plane ${\bf H}^2\subset{\bf H}^3$ (see \cite{Shch1},
proposition 1). Thus one can assume $\Gamma=\Or_{0}(1,2)$ and
$\Gamma_{0}=\{e\}$. Obviously, requirements of theorem \ref{th6}
are satisfied.
\begin{enumerate}
\item First consider the subcase $\beta_{0}\ne0$.
In this case according to theorem \ref{th6} the reduced phase
space $\widetilde{M}_{\beta_{0}}$ of the two-body problem is
diffeomorphic to the space
$$
T^{*}I\times\mathcal{O}_{\beta_{0}},
$$
and the reduced Hamilton function has the form
\begin{equation*}
\widetilde{H}_h=\frac{(1-r^2)^2}{8mR^2}p_r^2+\frac1ap_{4}^2 +
\frac12A_hp_{3}^{2}+\frac12C_hp_{5}^{2}+B_hp_{3}p_{5} + U(r).
\end{equation*}
Here $p_{3},p_{4},p_{5}$ are coordinates on the space
$\mathfrak{so}^{*}(1,2)$ and
$$
[p_{3},p_{4}]_{P}=p_{5},\,[p_{4},p_{5}]_{P}=-p_{3},\,[p_{5},p_{3}]_{P}=p_{4}.
$$
The orbit $\mathcal{O}_{\beta_{0}}$ is defined by the equation
$$
p_{3}^{2}-p_{4}^{2}-p_{5}^{2}=\mu.
$$
For $\mu>0$ the orbit $\mathcal{O}_{\beta_{0}}$ is a one sheet of
a two-sheet hyperboloid (diffeomorphic to the plane
$\mathbb{R}^{2}$), for $\mu=0$ it is the cone without vertex
(diffeomorphic to $\mathbb{R}^{2}\backslash\pt$), and for $\mu<0$
it is a one-sheet hyperboloid (diffeomorphic to the cylinder
$\mathbb{R}\times{\bf S^{1}}$).

\item  The last subcase $\beta_{0}=0$ corresponds to particle motion along a common geodesic
(see \cite{Shch1}, proposition 2).  Here
$\mathcal{O}_{\beta_{0}}=\pt$ and one gets
$$
\widetilde{M}_{0}=T^*I,\; \widetilde{H}_s =
\frac{(1-r^2)^2}{8mR^2}p_r^2+U(r).$$

\end{enumerate}
\end{enumerate}

\section{Conclusion}

In the present paper we have found the expression of the two-body
Hamiltonian on spaces ${\bf S}^n$ and ${\bf H}^n$ through a radial
differential operator and invariant differential operators on a
homogeneous spaces of isometry groups. This expression enables to
find some explicit series of energy levels for two particles on
the sphere ${\bf S}^3$. The most part of these series corresponds
to equal particle masses. Probably, the quasi-exactly solvability
of this quantum problem is connected with the existence of some
closed trajectories of the corresponding classical system.
Clearly, it is not difficult to find circular trajectories, when
the distance between particles with equal masses is constant.

A connection of closed trajectories of some non-integrable
classical mechanical problem with the spectrum of the
corresponding quantum mechanical system was studied in many papers
(see the overview and references in \cite{Gutz}). It would be
interesting to find such a connection in the problem under
consideration and also to calculate some exact spectral series for
the two-body problem on ${\bf S}^n,\,n\geqslant4$.

It was conjectured in \cite{Shch2} that the two-body Hamiltonian
on the hyperbolic plane ${\bf H}^2$ has no discrete energy levels.
The same seems to be valid for spaces ${\bf H}^n,\,n\geqslant3$.

The explicit form of the Hamilton function for the reduced
two-body problem in constant curvature spaces, founded in
\cite{Shch1} with a help of computer algebraic calculations, was
used there to prove the absence of particles collision. In the
present paper we have derived the explicit form of the reduced
Hamilton function without computer calculations and clarify its
connection with the two-body quantum Hamiltonian. This form of the
Hamiltonian reduction seems to be the most natural from the
geometric point of view, since the "radial" degree of freedom,
invariant w.r.t.\ the isometry group, is isolated as a direct
factor and another direct factor corresponds to the cotangent
bundle over a homogeneous manifold of the isometry group. The only
a priori integrable case of the reduced classical two-body problem
with a central interaction on constant curvature spaces, different
from particles movement along a common geodesic, corresponds to
the reduced Hamilton function (\ref{IntegrableCase}).

\end{document}